\documentclass
{ptephy_v1}

\usepackage{color}

\newcommand{\eqb}{\begin{equation}}
\newcommand{\eqe}{\end{equation}}
\newcommand{\eqbnon}{\begin{equation*}}
\newcommand{\eqenon}{\end{equation*}}

\newcommand{\eqab}{\begin{eqnarray}}
\newcommand{\eqae}{\end{eqnarray}}
\newcommand{\eqabnon}{\begin{eqnarray*}}
\newcommand{\eqaenon}{\end{eqnarray*}}

\newcommand{\seqb}{\begin{subequations}}
\newcommand{\seqe}{\end{subequations}}

\newcommand{\defeq}{:=}

\newcommand{\us}[1]{^{\rm #1}} 
\newcommand{\ls}[1]{_{\rm #1}} 

\newcommand{\pd}[2]{\dfrac{\partial #1}{\partial #2}}
\newcommand{\od}[2]{\dfrac{{\rm d} #1}{{\rm d} #2}}
\newcommand{\diff}[1]{{\rm d}#1}

\newcommand{\sgra}{Sgr\,A$^{\ast\,}$}
\newcommand{\Hu}{\mathcal{H}\ls{u}}

\newcommand{\ur}{u\ls{r}}
\newcommand{\Hk}{\mathcal{H}\ls{k}}

\newcommand{\kr}{k\ls{r}}

\newcommand{\ra}{{\rm RA}}
\newcommand{\dra}{\Delta Y}
\newcommand{\dec}{{\rm Dec}}
\newcommand{\ddec}{\Delta X}
\newcommand{\rv}{{\rm RV}}
\newcommand{\rs}{z_{\rm rs}}
\newcommand{\uncertain}[1]{\delta [#1]}

\newcommand{\ktilde}{\widetilde{k}}
\newcommand{\wtilde}{\widetilde{w}}
\newcommand{\ltilde}{\widetilde{l}_z}
\newcommand{\sigmatilde}{\widetilde{\sigma}}
\newcommand{\uzero}{^{\rm (0)}}
\newcommand{\lzero}{_{\rm (0)}}
\newcommand{\uone}{^{\rm (1)}}
\newcommand{\lone}{_{\rm (1)}}

\newcommand{\unit}[1]{\text{#1}}

\numberwithin{equation}{section}

\begin{document}

\title{
Parametrized-Post-Newtonian Test of Black Hole Spacetime for Galactic Center Massive Black Hole \sgra : 
Formulation and $\chi^2$ Fitting
}


\author{Hiromi Saida
}
\affil{
Department of Physics, 
Daido University, 
10-3, Takiharu-cho, Minami-ku, Nagoya, Aichi 457-8530, Japan
\email{saida@daido-it.ac.jp} 
}

\author{Sena A.\ Matsui}
\affil{
Division of Particle and Astrophysical Science, 
Nagoya University, 
Furo-cho, Chikusa-ku, Nagoya, Aichi 464-8602, Japan
}

\author[2,3]{Tsutomu T.\ Takeuchi}
\affil{
The Research Center for Statistical Machine Learning, 
the Institute of Statistical Mathematics, 
10-3 Midori-cho, Tachikawa, Tokyo 190-8562, Japan
}

\author{Shogo Nishiyama}
\affil{
Faculty of Education, 
Miyagi University of Education, 
149 Aramaki-aza-Aoba, Aoba-ku, Sendai, Miyagi 980-0845, Japan
}

\author[1]{Rio Saitou}

\author{Yohsuke Takamori}
\affil{
Department of Civil Engineering, 
National Institute of Technology, Wakayama College, 
77 Noshima, Nada-cho, Gobo, Wakayama 644-0023, Japan
}

\author{Masaaki Takahashi}
\affil{
Department of Physics and Astronomy,
Aichi University of Education, 
1 Hirosawa, Igaya-cho, Kariya, Aichi 448-8542, Japan
}



\begin{abstract}%
We have performed a parametrized post-Newtonian (PPN) test of a black hole spacetime using observational data of the star S0-2/S2 orbiting the massive black hole at our galactic center \sgra\!.
After introducing our PPN model of black hole spacetime, we report the result of $\chi^2$ fitting of the PPN model with the observational data.
A new finding through our PPN model is a detectability of the gravitational lens effect on the null geodesics connecting S0-2 and observer under the present observational uncertainties, if a PPN parameter is about one order larger than the value for general relativity case. 
On the other hand, the effect of black hole spin on the S0-2's motion is not detectable. 
Thus our present PPN test is performed with spherically symmetric vacuum black hole spacetime. 
The resultant value of the PPN parameter, which corresponds to the minimum $\chi^2$, implies that the gravitational field of \sgra is not of Schwarzschild metric or that
there exists a sufficient amount of dark matters around \sgra to be detected by present telescopes. 
However, the difference between the minimum $\chi^2$ and the $\chi^2$ of Schwarzschild case is not large enough to ensure a statistical significance of non-Schwarzschild result. 
A more precise statistical analysis than $\chi^2$ statistics is necessary to extract a statistically significant information of the gravitational field of \sgra from present observational data. 
We will report a result by a Bayesian analysis in next paper.
\end{abstract}

\subjectindex{E00, E03, E31, E36 \quad/\quad Accepted for publication at 21 Aug. 2024}

\maketitle

\section{Introduction}
\label{sec:intro}

Today, it is a common understanding that a massive black hole of mass $\simeq 4.0\times 10^6 M_\odot$ exists at the center of our galaxy, called Sagittarius A$^\ast$ (\sgra), where $M_\odot$ is the mass of sun. 
It is also found that more than a hundred of stars, called S-stars, are orbiting around \sgra.
These celestial objects at our galactic center are observed by radio and infrared telescopes, since optical photons are not useful because of the strong absorption by interstellar dusts and X-rays are not useful because of its low resolution with present X-ray telescopes. 
Stellar motions are observed with near infrared photons, and interstellar gases are observed with radio waves. 
The first strong evidences of the existence of massive black hole at a ``compact strong radio source'' \sgra were given by observing the motions of a few S-stars~\cite{ref:ghez+2000,ref:schoedel+2002}, but not based on general relativity because the observational uncertainties at that time were larger than general relativistic effects such as the orbital shift of S-stars and the gravitational redshift received by photons coming from S-stars.
In these first evidences, although no star's entire orbital period had been observed, the fitting of the observational data with Newtonian elliptical motions had revealed the high dense concentration of extremely large mass at \sgra whose spatial size was estimated less than 1~AU. 
(The mass was estimated to be $3.7\times 10^6 M_\odot$ at that time, while the present estimation is about $4.0\times 10^6 M_\odot$ as shown in Sect.\ref{sec:fitting}.) 
This extreme high density at \sgra was reasonably regarded as a massive black hole candidate.
Then the leaders of these pioneering observational studies~\cite{ref:ghez+2000,ref:schoedel+2002} were awarded the Nobel prize in physics 2020.

Most of known S-stars have long orbital periods of more than hundreds of years. 
Their distances to \sgra are so long that the observational uncertainties of the present instruments are larger than general relativistic effects appearing in the dynamics of those stars. 
Further, although a few S-stars, which were found in recent years, have short orbital periods of several years and close distances to \sgra, they are so faint that the observational uncertainties are larger than general relativistic effects in their dynamics. 
However, it is a good fortune that an S-star called S0-2/S2
has a rather short orbital period of $16$ years and a short distance to \sgra ranging from $120\,\unit{AU}$ to $1900\,\unit{AU}$, and is not so faint that the present largest infrared telescopes possess the capability for detecting a general relativistic effect in S0-2's dynamics. 
(Note that the name ``S0-2'' is given by American group's nomenclature~\cite{ref:ghez+2000}, while ``S2'' is by European group's nomenclature~\cite{ref:schoedel+2002}. 
We adopts the American group's one.)

The observational data of S0-2 have been obtained since 1992 by European group, 1995 by American group and 2014 by our Japanese group. 
The total of all data covers about two orbital periods of S0-2. 
The S0-2 passed the closest point to \sgra (the pericenter on its orbit) in 2002 and 2018. 
The developments of observational instruments, especially the adaptive optics systems between 2002 and 2018, enabled us to measure the general relativistic effects during the pericenter passage in 2018, while the observational uncertainties of data until 2017 are larger than the general relativistic effects. 
The general relativistic effects raised from the mass of \sgra have become detectable in 2018, but the effects of spin of \sgra do not. 
Further, it should be noted that no evidence of interstellar matters among \sgra and S-stars has been found so far, and that the distances between S-stars are so long that the gravitational interaction between them are negligible as estimated in Sect.\ref{subsec:ppn-system}. 
This fact means that the system of \sgra and S0-2 can be regarded as an isolated two body gravitational system. 
Therefore, using the observational data obtained since 1990s to 2018, the Newtonian elliptical motion of S0-2 and Schwarzschild's geodesic motion of S0-2 were compared. 
Then the Newtonian gravity was rejected~\cite{ref:gravity2018,ref:do+2019,ref:saida+2019}.
Now, the current interesting issue is the comparison of Schwarzschild metric with the other metrics.

In this paper, we introduce a parametrized-post-Newtonian (PPN) model of black hole spacetime. 
As explained in detail in Sect.\ref{sec:ppn}, the PPN metric is a modification of Kerr metric by introducing some artificial parameters which express how the PPN metric deviates from Kerr metric. 
Such artificial parameters are called the PPN parameters.
Let us denote the collection of all PPN parameters as ${\bf X}\ls{ppn}$, and the PPN metric as ${\bf g}({\bf X}\ls{ppn},m,a)$, where $m$ and $a$ are respectively the mass and spin angular momentum of black hole. 
Our aim is to determine the value of ${\bf X}\ls{ppn}$ by fitting the geodesic motion of S0-2 on ${\bf g}({\bf X}\ls{ppn},m,a)$ with the observational data of S0-2's motion. 
If the resultant value of ${\bf X}\ls{ppn}$ is not the value for Kerr metric, then the following two possibilities, or the hybrid of them, arise:
\begin{itemize}
\item[(I)]
If the region around \sgra is vacuum, then the general relativity is rejected for the gravity of \sgra.
\item[(II)]
If the general relativity is the correct theory of gravity, then the region around \sgra is not vacuum. 
There may exist dark matters whose amount is much enough to be detectable by the present telescopes through the motion of S0-2.
\end{itemize}

Note that a new finding through our PPN model, as shown in Sect.\ref{subsec:obs-PNPM}, is a detectability of the gravitational lens effect on the null geodesics connecting S0-2 and observer under the present observational uncertainties, if a PPN parameter is about one order larger than the value for general relativity case. 
This detectability of lens effect had not been recognized in all previous papers of all groups observing S0-2~\cite{ref:ghez+2000,ref:schoedel+2002,ref:gillessen+2017,
ref:gravity2018,ref:do+2019,ref:saida+2019,ref:gravity2020}. 
On the other hand, the effect of black hole spin on the S0-2's motion is not detectable. 
Thus, our present PPN test is performed with spherically symmetric vacuum black hole spacetime with taking into account the gravitational lens effect.

We try to search for the best-fitting value of ${\bf X}\ls{ppn}$ by a simple statistical method, the $\chi^2$ fitting.
If $\chi^2$ fitting gives us a statistically significant discrimination between the PPN case and the Schwarzschild case, then it is an enough information for discussing the physics of gravity around \sgra. 
However, if we can not obtain a statistically significant result, then the need for a more complicated method such as a Bayesian analysis arises.

Here we should note that all parameters in the PPN formalism, not only ${\bf X}\ls{ppn}$ but also all the other parameters such as $\{m,a\}$ and the initial condition of S0-2's motion, should be evaluated by fitting our PPN model with the present observational data. 
It is not good to search for best-fitting values of only ${\bf X}\ls{ppn}$ with fixing the other parameters at the Schwarzschild case, because the global minimum of $\chi^2$ may be out of the best-fitting values of the Schwarzschild case. 
Concerning this issue, a previous work using a PPN model~\cite{ref:gainutdinov2020} is not  statistically sufficient, because the parameters such as the mass of black hole are fixed at the Schwarzschild case. 
In addition, the previous work~\cite{ref:gainutdinov2020} do not aware of the possible detectability of gravitational lens effects. 
Then, the resultant values of PPN parameters in the previous work~\cite{ref:gainutdinov2020} are largely different from our result shown in Sect.\ref{sec:fitting}.
Hence, our analysis is the first consistent application of PPN model to the system of \sgra and S0-2.

In Sect.\ref{sec:ppn}, our PPN model of black hole spacetime is introduced. 
Further, although the present data do not enable us to test the spin effects of black hole, we develop our PPN model so as to include the largest spin effects, because those effects are expected to be measurable with the near future large telescope (e.g. Thirty-Meter-Telescope). 
Sect.\ref{sec:obs} is for deriving the formulas of observational quantities. 
The need for the gravitational lens effects for null geodesics, which were ignored in previous works, is also discussed under the present observational uncertainties. 
Sect.\ref{sec:fitting} is devoted to the $\chi^2$ fitting of the PPN model prediction with the observational data taken by European, American and our Japanese groups, and the best-fitting parameter values are shown as well. 
Sect.\ref{sec:discussions} is for discussions.

The units used throughout this paper are of $c=1$ and $G=1$. 
When showing the numerical values of physical quantities, the constants $c$ and $G$ will be shown explicitly. 
Greek index $\mu = 0, 1, 2, 3$ denotes the spacetime components of tensors, and Latin index $j = 1, 2, 3$ denotes the spatial components of tensors.

\section{Parameterized post-Newtonian/Minkowskian formulation of our problem}
\label{sec:ppn}

After introducing the parameter of post-Newtonian expansion in Sect.\ref{subsec:ppn-system}, the PPN formulations of metric, timelike geodesics and null geodesics are derived successively in Sect.\ref{subsec:ppn-metric}, \ref{subsec:ppn-timelike}, and \ref{subsec:ppn-null}. 
In Sect.\ref{subsec:ppn-setup}, we clarify the coordinate system for observation, the setup of the initial conditions of S0-2's motion, and all the parameters which are to be evaluated by fitting our PPN model with observational data.

\subsection{Post-Newtonian expansion parameter for the system of \sgra and S0-2}
\label{subsec:ppn-system}

The parameter $\varepsilon$ of post-Newtonian (PN) expansion for the system composed of \sgra and S0-2 is defined by
\eqb
\label{eq:ppn-pnparameter}
 \varepsilon(r) \defeq \dfrac{GM\ls{BH}}{c^2 r} \simeq \dfrac{v(r)^2}{2 c^2} \,,
\eqe
where $r$ is the distance (radial coordinate) of S0-2 to \sgra, and $v(r)$ is the speed of S0-2 at $r$.  
This $\varepsilon(r)$ can be interpreted as the specific potential energy of S0-2.
The similar equality ``$\simeq$'' in Eq.\eqref{eq:ppn-pnparameter} is due to a general fact that the potential energy and the kinetic energy have the values of similar order for the object moving on a bounded orbit around a central mass.

Although the precise best-fitting values of parameters such as the mass of \sgra are derived later in Sect.\ref{sec:fitting}, approximate values of those parameters have already been known~\cite{ref:do+2019,ref:saida+2019,ref:gravity2020}. 
The approximate values of parameters needed in this section are as follows.
\eqb
\label{eq:ppn-value}
\begin{split}
 \text{Mass of \sgra} &: M\ls{BH} \sim 4 \times 10^6 M_\odot
\\
 \text{Distance from Sun to \sgra} &: R\ls{GC} \sim 8 \,\unit{kpc}
\\
 \text{Pericenter distance of S0-2 to \sgra} &: r\ls{peri} \sim 120 \,\unit{AU}
 \,,
\end{split}
\eqe
where the suffix ``GC'' means the galactic center. 
When the PN parameter $\varepsilon$ is evaluated at the pericenter of S0-2, its value is
\eqb
\label{eq:ppn-pnparameter-peri}
 \varepsilon\ls{peri} \defeq 
 \varepsilon(r\ls{peri}) = 
 \dfrac{GM\ls{BH}}{c^2 r\ls{peri}} \sim 10^{-3} \,,
\eqe
while the PN parameter evaluated at the surface of the sun is $\varepsilon_\odot \defeq GM_\odot/(c^2 r_\odot) \sim 10^{-6}$, where $r_\odot$ is the radius of the sun. 
The gravity produced by \sgra at S0-2's pericenter is about three orders of magnitude stronger than the gravity at the surface of sun.

The PN parameter of the gravity between S0-2 and one of the other S-stars is roughly estimated as 
$\varepsilon\ls{S} \defeq GM_\odot/(c^2 r\ls{peri}) = 
10^{-6} \varepsilon\ls{peri} \sim 10^{-9}$, 
where the mass of each star and the distance among S-stars are respectively approximated by the solar mass $M_\odot$ and the S0-2's pericenter distance $r\ls{peri}$. 
The gravitational effect by $\varepsilon\ls{S} \sim 10^{-9}$ is so small that the present telescopes cannot detect. 
Further, because significant interstellar gases around S-stars are not found, the so-called dynamical friction on S0-2 can be neglected. 
The effects of stellar spin of S0-2 and stellar wind from S0-2 are also negligible. 
Hence, we assume that the dynamics of S0-2 and photons emitted by S0-2 are described by, respectively, a timelike geodesic and null geodesics on the gravitational field produced by \sgra.

Further, the telescopes detect the photons coming from S0-2, and any observable quantity is read from the detected photons. 
Therefore, we need the PPN formulations of the metric tensor of \sgra, the timelike geodesic of S0-2 and the null geodesics of photons connecting S0-2 and the observer. 
These PPN formulations are given in the following subsections.

\subsection{Parameterized post-Newtonian expansion of black hole metric}
\label{subsec:ppn-metric}

We develop the PPN formulas starting from the Kerr metric $g_{\mu' \nu'}\us{(Kerr)}$ in the Boyer-Lindquist coordinates, $x^{\mu'} = (t,r,\theta,\varphi)$,
\seqb
\label{eq:kerr}
\eqb
 \diff{s}^2
 = g\us{(Kerr)}_{\mu' \nu'} \diff{x}^{\mu'} \diff{x}^{\nu'}
 =
 - \dfrac{\Sigma \Psi}{Z} \,\diff{t}^2
 + \dfrac{Z}{\Sigma}\sin^2\theta\,\Bigl[ \omega \,\diff{t} - \diff{\varphi} \Bigr]^2
 + \dfrac{\Sigma}{\Psi} \,\diff{r}^2 + \Sigma \,\diff{\theta}^2
 \,,
\eqe
where the metric functions are
\eqb
\label{eq:kerr-function}
\begin{split}
 \Psi(r) &\defeq r^2 + a^2 - 2m r
\\
 \Sigma(r,\theta) &\defeq r^2 + a^2 \cos^2\theta
\\
 Z(r,\theta) &\defeq (r^2+a^2) \Sigma(r,\theta) + 2 m r a^2 \sin^2\theta
 \,=\, (r^2+a^2)^2 - \Psi(r) a^2 \sin^4\theta
\\
 \omega(r,\theta) &\defeq \dfrac{2 m r}{Z(r,\theta)}\,a
 \,,
\end{split}
\eqe
\seqe
and $m$ and $a$ are respectively the black hole's mass and spin angular momentum in the length dimension. 
We transform from the spherical coordinates $x^{\mu'} = (t,r,\theta,\varphi)$ to the Cartesian-like coordinates $x^\mu = (t,x,y,z)$,
\eqb
\label{eq:cartesian-spherical}
 x = r\sin\theta \cos\varphi
 \quad,\quad
 y = r\sin\theta \sin\varphi
 \quad,\quad
 z = r\cos\theta \,,
\eqe
where $z$-axis is the spin axis of Kerr black hole. 
Using the Cartesian-like coordinates, we avoid some numerical difficulties in the spherical coordinates arising from $\theta =0$ and $\pi$.

The PN parameter \eqref{eq:ppn-pnparameter} is expressed as $\varepsilon(r) = m/r$ in Eqs.\eqref{eq:kerr}. 
Let us expand the Kerr metric $g_{\mu\nu}\us{(Kerr)}$ with $\varepsilon(r)$ in the Cartesian-like coordinates, and introduce artificial parameters at each term needed for later discussions. 
The metric $g_{\mu\nu}$ obtained by this procedure is
\eqb
\label{eq:ppn-all}
\begin{split}
 g_{00} &=
 -1 + 2N\ls{t}\, \varepsilon(r) + A \varepsilon(r)^2 + O(\varepsilon^3)
\\
 g_{01} &=
 N\ls{s}\, \dfrac{x}{r} \varepsilon(r)
 + 2 C_\perp \dfrac{a}{m}\, \dfrac{y}{r} \varepsilon(r)^2 + O(\varepsilon^3)
\\
 g_{02} &=
 N\ls{s}\, \dfrac{y}{r} \varepsilon(r)
 - 2 C_\perp \dfrac{a}{m}\, \dfrac{x}{r} \varepsilon(r)^2 + O(\varepsilon^3)
\\
 g_{03} &=
 N\ls{s}\, \dfrac{z}{r} \varepsilon(r)
 + 2 C_z \dfrac{a}{m}\, \dfrac{z}{r} \varepsilon(r)^2 + O(\varepsilon^3)
\\
 g_{ij} &=
 \delta_{ij} + 2 B \dfrac{x^i x^j}{r^2}\varepsilon(r) + O(\varepsilon^2)
  \,,
\end{split}
\eqe
where $A$, $B$, $C_z$, $C_\perp$ $N\ls{t}$ and $N\ls{s}$ are the non-dimensional artificial parameters under the assumption of the stationary axisymmetry about $z$-axis and the spherical symmetry for the non-spinning case $a=0$. 
This expansion~\eqref{eq:ppn-all} is our \emph{parametrized post-Newtonian (PPN) expansion} of the Kerr metric in the Cartesian-like coordinates, and the coefficient parameters 
${\bf X}\ls{ppn} = \{A,B,C_z,C_\perp,N\ls{t},N\ls{s}\}$ 
are the \emph{PPN parameters}. 
The value of ${\bf X}\ls{ppn}$ corresponding to the post-Newtonian expansion of Kerr metric \eqref{eq:kerr} is 
\eqb
\label{eq:ppn-kerrcase}
 \{A,B,C_z,C_\perp,N\ls{t},N\ls{s}\}\us{(Kerr)} = \{0,1,0,1,1,0\} \,.
\eqe

Here let us note about the spin parameter $a$. 
We consider that, if a condition of extremely high spin parameter $|a/m| \gg 1$ was satisfied, then the Newtonian fitting of past data of S-stars in 2000's (see the first paragraph in Sect.\ref{sec:intro}) could not make a statistically significant result. 
Therefore, we assume the spin parameter satisfying
\eqb
\label{eq:ppn-spin}
 O\Bigl(\dfrac{a}{m}\Bigr) \lesssim 1
 \,.
\eqe
This means that we do not necessarily restrict our analysis to a slow spin case, but include the high spin case of $O(a/m) \sim 1$. 
Indeed the PPN expansion \eqref{eq:ppn-all} is based on the expansion of Kerr metric by only $\varepsilon(r)$, and no expansion by $a/m$ is introduced in Eq.\eqref{eq:ppn-all}.
\footnote{
Because only the "1st order" spin parameter $a/m$ appears in $g_{0j}$ of Eq.\eqref{eq:ppn-all}, one might think the expansion by $a/m$ was also introduced. 
However, the appearance of $a/m$ in Eq.\eqref{eq:ppn-all} is due to the metric function $\omega(r,\theta)$ in Kerr metric's $g_{0j}\us{(Kerr)}$ components, and the coefficients of terms of $O(\varepsilon^3)$ in Eq.\eqref{eq:ppn-all} include $(a/m)^2$ and $a/m$ due to the metric functions in Eq.\eqref{eq:kerr-function}. 
}

Next, the other note we need to clarify is the independent PPN parameters under the aim of this paper. 
Although there appear six PPN parameters in Eq.\eqref{eq:ppn-all}, three of them $\{C_\perp,N\ls{t},N\ls{s}\}$ are fixed to be the values of Kerr case \eqref{eq:ppn-kerrcase} as explained below.

On the parameter $N\ls{t}$, let us note that the term of $O(\varepsilon)$ in $g_{00}$ expresses the Newtonian gravity, as will be shown by the PPN expansion of timelike geodesics in Sect.\ref{subsec:ppn-timelike} and Appendix~\ref{app:ppn}. 
Hence, by requiring that the Newtonian potential $m/r (= \varepsilon)$ is recovered at non-relativistic situations, we fix as $N\ls{t} = 1$. 
Further, one benefit of this fixation is the resolution of a degeneracy between ${\bf X}\ls{ppn}$ and $m$. 
From Eq.\eqref{eq:ppn-all} one can understand that one of the six PPN parameters in ${\bf X}\ls{ppn}$ cannot be distinguished from the mass $m$ by observations of S0-2, because not only ${\bf X}\ls{ppn}$ but also $m$ are to be evaluated by fitting with observational data. 
By the requirement $N\ls{t} = 1$, the other PPN parameters are distinguished from $m$.

On the parameter $N\ls{s}$, let us note that the terms of $O(\varepsilon)$ in $g_{0j}$ raise a relativistic effect (so-called 0.5PN effect) which is larger than the pericenter shift of S0-2 (so-called 1PN effects), as will be shown by the PPN expansion of timelike geodesics in Sect.\ref{subsec:ppn-timelike} and Appendix~\ref{app:ppn}. 
This is interpreted as a modification of Newtonian potential so that the potential depends on the velocity of S0-2. 
If such velocity dependence in Newtonian potential exists, it should have to be already found so far through the observations of S0-2. 
However, such an effect has not been found. 
Therefore we fix as $N\ls{s} = 0$.

On the parameter $C_\perp$, one can understand from Eq.\eqref{eq:ppn-all} that the PPN parameters $C_\perp$ or $C_z$ cannot be distinguished from the spin $a$ by observations of S0-2, because not only $\{C_\perp,C_z\}$ but also $a$ are to be evaluated by fitting with observational data. 
Therefore, we fix as $C_\perp = 1$, and leave $C_z$ free.

From the above, the form of PPN metric we are going to use in the following sections is
\seqb
\label{eq:ppn-1PN1PM}
\eqb
\begin{split}
 g_{00} &=
 -1 + 2\, \varepsilon(r) + A \varepsilon(r)^2 + O(\varepsilon^3)
\\
 g_{0j} &=
 2 \dfrac{a}{m}\, D_j \, \varepsilon(r)^2 + O(\varepsilon^3)
\\
 g_{ij} &=
 \delta_{ij} + 2 B \dfrac{x^i x^j}{r^2}\varepsilon(r) + O(\varepsilon^2)
  \,.
\end{split}
\eqe
where $D_j \defeq (y/r , -x/r , C_z z/r)$. 
For the later use, let us show the inverse metric $g^{\mu\nu}$,
\eqb
\label{eq:ppn-1PN1PM-inverse}
\begin{split}
 g^{00} &=
 -1 - 2 \varepsilon(r) - (4+A) \varepsilon(t)^2 + O(\varepsilon^3)
\\
 g^{0j} &=
 2 \dfrac{a}{m}\, D^j \, \varepsilon(r)^2 + O(\varepsilon^3)
\\
 g^{ij} &=
 \delta^{ij} - 2 B \dfrac{x^i x^j}{r^2}\varepsilon(r) + O(\varepsilon^2)
  \,,
\end{split}
\eqe
\seqe
where $\delta_{ij} = \delta^{ij}$ is the Kronecker's delta, and $D^j = D_j$. 
Note that the terms of $O(\varepsilon^2)$ in $g_{0j}$ express the largest effect of the black hole spin. 
This spin effect is not detectable by the present telescopes as will be shown in Sect.\ref{subsec:obs-PNPM}.
However we derive our formulas without eliminating those terms in this section, because the largest spin effect is expected to be detectable by the near future telescopes, for example the Thirty-Meter-Telescope which is to be established in the Maunakea observatories.

Finally in this subsection, let us clarify the relation between our PPN metric \eqref{eq:ppn-1PN1PM} and the so-called \emph{standard PPN gauge} established by C.M.Will~\cite{ref:will1993}. 
The standard PPN gauge is originally formulated for self-gravitating fluid systems up to the terms of the order next to Newtonian gravity. 
In this gauge, the spatial coordinates are fixed so that the spatial parts of metric components are proportional to $\delta_{ij}$ up to $O(\varepsilon)$. 
Therefore, the transformation between our Cartesian-like coordinates $x^\mu = (t,x,y,z)$ and the standard PPN coordinates $\bar{x}^{\bar{\mu}} = (\bar{t},\bar{x},\bar{y},\bar{z})$ are given by
\seqb
\label{eq:ppn-standard}
\eqb
 t = \bar{t} \quad,\quad
 x^j = \left(1+B \dfrac{m}{\bar{r}}\right) \bar{x}^{\bar{j}} \,,
\eqe
where $\bar{r} \defeq \sqrt{\bar{x}^2+\bar{y}^2+\bar{z}^2}$. 
The metric components in this coordinates are
\eqb
\begin{split}
 g_{\bar{0}\bar{0}} &=
 -1 + 2\, \bar{\varepsilon}(\bar{r}) + (A-2B)\, \bar{\varepsilon}(\bar{r})^2 + O(\bar{\varepsilon}^3)
\\
 g_{\bar{0}\bar{j}} &=
 2 \dfrac{a}{m}\, \bar{D}_{\bar{j}}\, \bar{\varepsilon}(\bar{r})^2 + O(\bar{\varepsilon}^3)
\\
 g_{\bar{i}\bar{j}} &=
 \left( 1 + 2 B \bar{\varepsilon}(\bar{r}) \right)\, \delta_{\bar{i}\bar{j}} + O(\varepsilon^2)
  \,,
\end{split}
\eqe
\seqe
where $\bar{\varepsilon}(\bar{r}) \defeq m/\bar{r}$ and $\bar{D}_{\bar{j}} \defeq (\bar{y}/\bar{r} , -\bar{x}/\bar{r} , C_z \bar{z}/\bar{r})$.

\subsection{Parameterized post-Newtonian expansion of timelike geodesics}
\label{subsec:ppn-timelike}

Before proceeding to the formulation of PPN expansion of timelike geodesics, let us point out one problem in solving numerically the geodesic equations of Kerr metric. 
In the Boyer-Lindquist coordinates, the usual form of the geodesic equations $u^\nu \nabla_\nu u^\mu = 0$ give a second order differential equation of the radial coordinate $r(\lambda)$ of the geodesic,
\eqb
\label{eq:ppn-problem}
 \od{^2 r(\lambda)}{\lambda^2} =
 \pm \sqrt{\text{combination of the metric functions \eqref{eq:kerr-function}}}
 \,,
\eqe
where $\lambda$ is an affine parameter, and the geodesic equation of the angular coordinate $\theta(\lambda)$ of the geodesic has the same structure. 
The problem in numerical calculation arises from the signature ``$\pm$'' of the right-hand side. 
In calculating numerical integrations, the signature of the right-hand side should be specified. 
Once the signature is mistaken, a serious numerical error occurs. 
Especially in the vicinity of zeros of the right-hand side, the numerical code for the choice of the signature needs a special care. 
This problem is not removed in the Cartesian-like coordinates. 
Because the PPN expansion of the geodesic equations is essentially the expansion of the right-hand side of Eq.\eqref{eq:ppn-problem} by the PN parameter $\varepsilon(r)$, the problem of the signature is not removed.

However, let us note that this problem becomes manifest in the case that the geodesic equations are expressed as the second order differential equations of the coordinates $x^\mu(\tau)$. 
This problem can be removed in the Hamiltonian formalism of geodesic equations, in which the geodesic equations are formulated as the simultaneous first order differential equations of not only coordinates $x^\mu(\lambda)$ but also tangent 1-forms $u_\mu(\lambda)$. 
Therefore, we adopt the Hamiltonian formalism of geodesic equations.

\subsubsection{Hamiltonian}
\label{subsubsec:ppn-timelike-hamiltonian}

The dynamical variables in the Hamiltonian formalism of geodesic equations are the spacetime point on the geodesic $x^\mu(\tau)$ and the 1-form conjugate to the four velocity of the geodesic $u_\mu(\tau)$, where $\tau$ is the proper length along the geodesic. 
The 1-form $u_\mu(\tau)$ has no dimension, while the point $x^\mu(\tau)$ has the length dimension. 
For these dynamical variables, the Hamiltonian of geodesics is given by
\seqb
\label{eq:ppn-hamiltonian-timelike}
\eqb
\label{eq:ppn-Hu}
 \Hu(x,u) = \dfrac{1}{2} g^{\mu\nu}(x) u_\mu u_\nu \,,
\eqe
where $x$ and $u$ denote symbolically the dynamical variables, and the normalization constraint of four velocity is assigned for timelike geodesics,
\eqb
\label{eq:ppn-cosntraint-timelike}
 \Hu(x,u) = -\dfrac{1}{2}
 \,.
\eqe
\seqe
The Hamilton equations are given by
\eqb
\label{eq:ppn-hamiltoneq-timelike}
 \od{u_\mu(\tau)}{\tau} = -\pd{\Hu(x,u)}{x^\mu} \quad,\quad
 \od{x^\mu(\tau)}{\tau} = \pd{\Hu(x,u)}{u_\mu}
 \,.
\eqe
The solution of these equations under the constraint \eqref{eq:ppn-cosntraint-timelike} is the timelike geodesic. 
The Hamilton equations \eqref{eq:ppn-hamiltoneq-timelike} are the first order differential equations. 
We construct our PPN formulations of timelike geodesics from the Hamiltonian \eqref{eq:ppn-Hu}.
\footnote{
Following the ordinary procedure of the analytical mechanics, the Lagrangian is related to the Hamiltonian through the Legendre transformation of the dynamical variables, $L(x,\dot{x}) = u_\mu \dot{x}^\mu - H$, where $\dot{x} = \diff{x}/\diff{\tau}$. 
The Euler-Lagrange equations of this Lagrangian are the second order differential equations of the same form with $u^\nu \nabla_\nu u^\mu = 0$. 
}

Due to the stationary axial symmetry of spacetime, there are two conserved quantities along timelike geodesics,
\eqb
\label{eq:ppn-constant-timelike}
 E \defeq -u_0 = \text{constant} \quad,\quad
 L_z \defeq x u_2 - y u_1 = \text{constant}
 \,,
\eqe
where $\diff{E}/\diff{\tau} = 0$ and $\diff{L_z}/\diff{\tau} = 0$ are shown from $\partial\Hu/\partial t = 0$, $\partial\Hu/\partial\varphi = 0$ and Eqs.\eqref{eq:ppn-hamiltoneq-timelike}.
Physical meanings of $E$ and $L_z$ are respectively the energy and the angular momentum around z-axis of a test particle (the star S0-2) moving on the timelike geodesic, where $E$ has no dimension normalized by the mass energy of S0-2 and $L_z$ has the length dimension.

Note that, for Kerr spacetime, there exists the third constant of geodesic motions, Carter constant, due to the so-called hidden symmetry of spacetime described by Killing tensor. 
On the other hand, as shown by C.M.Will~\cite{ref:will2009} in Newtonian gravity, a special case of stationary axisymmetric Newtonian gravitational potential allows the existence of a ``Carter-like'' constant for motions of test particles, which is different from the energy and angular momentum. 
Hence, in our PPN model \eqref{eq:ppn-1PN1PM} which possesses the stationary axisymmetry, there may exist a special set of values of PPN parameters ${\bf X}\ls{ppn}$, other than the Kerr case \eqref{eq:ppn-kerrcase}, which generates a ``Carter-like'' constant for geodesic motions. 
However, even if such a special case exists in our PPN model, we do not fix the value of ${\bf X}\ls{ppn}$ at the special case, because our aim in this paper is the search of the value of ${\bf X}\ls{ppn}$ best-fitting with observational data. 
Hence, in this paper, we do not expect the existence of a third constant of geodesic motions. 
(The search for a ``Carter-like'' constant in our PPN model is an interesting issue, but not in the scope of this paper.)

The PPN expansion of $\Hu(x,u)$ is obtained by substituting the metric \eqref{eq:ppn-1PN1PM} into Eq.\eqref{eq:ppn-Hu},
\seqb
\label{eq:ppn-Hu-expansion}
\eqb
\begin{array}{rcll}
 \Hu(x,u)
 &=&
 \displaystyle
 \dfrac{1}{2}g^{00} E^2
 - g^{0j}\, u_j \, E
 + \dfrac{1}{2} g^{ij} u_i u_j
 &
\\[3mm]
 &=&
 \displaystyle
 - \dfrac{1}{2} E^2 + \dfrac{1}{2} \sum_{j=1}^3 u_j^2
 - E^2 \varepsilon(r)
 &\cdots\text{0PN : up to $O(\varepsilon)$}
\\[4mm]
 &&
 - B u\ls{r}^2 \varepsilon(r)
 - \dfrac{4 + A}{2} E^2 \varepsilon(r)^2
 &\cdots\text{1PN : $O(\varepsilon^2) = O(\varepsilon u^2)$}
\\[3mm]
 &&
 + 2 \dfrac{a}{m} \Bigl(\,
   \dfrac{L_z}{r} - C_z \dfrac{z}{r} u\ls{r}
   \,\Bigr)
   \,E\,\varepsilon(r)^2
 &\cdots\text{1.5PN : $O(\varepsilon^2 u) = O(\varepsilon^{2.5})$}
\\[3mm]
 &&
 + O(\varepsilon^3)
 &\cdots\text{higher PN}
 \,,
\end{array}
\eqe
where the summation of $i$ and $j$ by the Einstein rule is for spatial components, the order of terms is counted under the relation $O(u) = O(\varepsilon^{1/2})$ shown in Eq.\eqref{eq:ppn-pnparameter}, and $u\ls{r}$ is given by
\eqb
\label{eq:ppn-ur}
 \ur \defeq \dfrac{1}{r} x^j u_j \,.
\eqe
\seqe
In Eq.\eqref{eq:ppn-Hu-expansion}, ``$n$PN'' means the terms of $O(\varepsilon^{n+1})$. 
The 0PN terms express the Newtonian gravity in the framework of the special relativity. 
The 1PN terms express the largest non-Newtonian effect, which depends on the mass $m$ but not on the spin $a$. 
The 1.5PN terms express the largest effect depending on the spin $a$. 
Note that Eq.\eqref{eq:app-Hu-expansion} in Appendix~\ref{app:ppn} shows the PPN expansion of $\Hu$ with retaining $\{ N\ls{t} , N\ls{s} , C_\perp \}$ in the metric \eqref{eq:ppn-all}.

\subsubsection{Geodesic equations}
\label{subsubsec:ppn-timelike-eq}

The PPN timelike geodesic equations are obtained by substituting Eqs.\eqref{eq:ppn-Hu-expansion} into Eq.\eqref{eq:ppn-hamiltoneq-timelike}. 
This procedure, usually, gives the geodesic equations in which the dynamical variables are regarded as the functions of the proper time $\tau$. 
On the other hand, the observational data produce, for example, the position of S0-2 $x(t)$ as the function of the observer's proper time which corresponds to the coordinate time $t$, because the observer is far from \sgra. 
Therefore we formulate our PPN geodesic equations with regarding the dynamical variables as the functions of $t$ through the following transformation,
\eqb
\label{eq:ppn-timetransformation}
 \od{x^j(t)}{t} = \od{x^j(\tau)}{\tau}\left(\od{t(\tau)}{\tau}\right)^{-1} \quad,\quad
 \od{u_j(t)}{t} = \od{u_\mu(\tau)}{\tau}\left(\od{t(\tau)}{\tau}\right)^{-1} \,,
\eqe
where the right-hand sides are given by the Hamilton equations with regarding $\{ x^\mu(\tau) , u_j(\tau) \}$ as the functions of $\tau$, and $u_0$ is omitted because $u_0 = -E$ is constant. 
The PPN timelike geodesic equations through the transformation \eqref{eq:ppn-timetransformation} are as follows.
\seqb
\label{eq:ppn-timelikegeodesic}
\eqb
\label{eq:ppn-timelikegeodesic-xj}
\begin{array}{rcll}
 E \od{x^j(t)}{t}
 &=& E \od{x^j(\tau)}{\tau}\,\Bigl(\od{t(\tau)}{\tau}\Bigr)^{-1}
 &
\\
 &=&
 u_j
 &
 \cdots\text{0PN}
\\[0mm]
 &&
 - 2 B\, \ur \dfrac{x^j}{r} \,\varepsilon(r)
 - 2 u_j \,\varepsilon(r)
 &
 \cdots\text{1PN}
\\[3mm]
 &&
 +2 \dfrac{a}{m}
 \Bigl(\, \dfrac{d^j}{r} - C_z \dfrac{z x^j}{r^2}
 \,\Bigr) E^2 \varepsilon(r)^2
 &
 \cdots\text{1.5PN}
\\[4mm]
 && 
 + O(\varepsilon^{2.5})
 &
 \cdots\text{higher PN}
\end{array}
\eqe
\eqb
\label{eq:ppn-timelikegeodesic-uj}
\begin{array}{rcll}
 E \od{u_j(t)}{t}
 &=&
 E \od{u_j(\tau)}{\tau}\,\Bigl(\od{t(\tau)}{\tau}\Bigr)^{-1}
 &
\\[2mm]
 &=&
 - E^2 \, \dfrac{x^j}{r^2}\,\varepsilon(r)
 &
 \cdots\text{0PN}
\\[2mm]
 &&
 \!\!\!
 \left.
 \begin{array}{l}
 + B\,\Bigl( 3 \dfrac{x^j}{r} \ur^2 - 2 u_j \ur \,\Bigr)\,\dfrac{\varepsilon(r)}{r}
 \\
 + \bigl( 6 - A \bigr) E^2 \dfrac{x^j}{r^2} \varepsilon(r)^2
 \end{array}
 \qquad\right\}
 &
 \cdots\text{1PN}
\\[7mm]
 &&
  + \Bigl( 6 P(x,u)\,\dfrac{x^j}{r} - 2 Q_j(x,u) \Bigr)\,E\,\dfrac{\varepsilon(r)^2}{r}
 &
 \cdots\text{1.5PN}
\\[3mm]
 && 
 + O(\varepsilon^3)
 &
 \cdots\text{higher PN}
 \,,
\end{array}
\eqe
where $u\ls{r}$ is given in Eq.\eqref{eq:ppn-ur}, $d^j = (-y,x,0)$ in 1.5PN terms of Eq.\eqref{eq:ppn-timelikegeodesic-xj}, and $P(x,u)$ and $Q_j(x,u)$ in 1.5PN terms of Eq.\eqref{eq:ppn-timelikegeodesic-uj} are
\eqb
\label{eq:ppn-PQ}
\begin{split}
 P(x,u) &\defeq
 - \dfrac{a}{m} \Bigl(\,
 \dfrac{L_z}{r} - C_z \dfrac{z}{r} \ur \,\Bigr)
\\
 Q_j(x,u) &\defeq
 - \dfrac{a}{m} \Bigl[\, q_j
 -C_z \Bigl\{
    \delta_{j3} \ur
    + \dfrac{z}{r}\Bigl(\, \dfrac{u_j}{r} - \dfrac{x^j}{r^2} \ur \,\Bigr)
 \,\Bigr\}\,\Bigr]
 \,,
\end{split}
\eqe
\seqe
where $q_j = (u_2 , -u_1 , 0)$. 
Concerning these geodesic equations, let us make two notes.
\begin{itemize}
\item
Newtonian gravity is recovered by focusing on 0PN terms in Eqs.\eqref{eq:ppn-timelikegeodesic}, where 3D velocity in Newtonian mechanics is given by $v\ls{Newton}^j \defeq u_j/(-u_0) = u_j/E$. 
\item
Although six dynamical variables $\{ x^j(t) , u_j(t) \}$ appear in Eqs.\eqref{eq:ppn-timelikegeodesic}, one of four variables $\{x(t),y(t),u_1(t),u_2(t)\}$ is dependent due to the conserved quantity $L_z$ in Eq.\eqref{eq:ppn-constant-timelike}. 
In integrating Eqs.\eqref{eq:ppn-timelikegeodesic} numerically, five dynamical variables need to be solved, once the values of $E$ and $L_z$ are specified through the initial conditions of S0-2 (see Sect.\ref{subsec:ppn-setup}). 
\end{itemize}

\subsection{Parameterized post-Minkowskian expansion of null geodesics}
\label{subsec:ppn-null}

Although the parameter $\varepsilon(r)$ in Eq.\eqref{eq:ppn-pnparameter} is called the ``post-Newtonian'' parameter, the expansion of null geodesics using $\varepsilon(r)$ as the expansion parameter is called the \emph{post-Minkowskian (PM)} expansion, because the leading order terms express null geodesics on Minkowski metric.

\subsubsection{Hamiltonian}
\label{subsubsec:ppn-null-hamiltonian}

As for timelike geodesics, we adopt the Hamiltonian formalism for null geodesics. 
The dynamical variables are the spacetime point on the geodesic $x^\mu(\sigma)$ and the tangent 1-form of the geodesic $k_\mu(\sigma)$, where $\sigma$ is an affine parameter along the geodesic. 
Hereafter, let $\sigma$ have the length dimension, and the 1-form $k_\mu(\sigma)$ has no dimension, while the point $x^\mu(\sigma)$ has the length dimension. 
For these dynamical variables, the Hamiltonian and the null condition are, respectively, given by
\seqb
\label{eq:ppn-hamiltonian-null}
\eqab
\label{eq:ppn-Hk}
 \Hk(x,k) &=& \dfrac{1}{2} g^{\mu\nu}(x) k_\mu k_\nu
\\
\label{eq:ppn-cosntraint-null}
 \Hk(x,k) &=& 0
 \,,
\eqae
\seqe
where $x$ and $k$ denote symbolically the dynamical variables. 
The Hamilton equations are given by
\eqb
\label{eq:ppn-hamiltoneq-null}
 \od{k_\mu(\sigma)}{\sigma} = -\pd{\Hk(x,k)}{x^\mu} \quad,\quad
 \od{x^\mu(\sigma)}{\sigma} = \pd{\Hk(x,k)}{k_\mu}
 \,.
\eqe
The solution of these equations under the constraint \eqref{eq:ppn-cosntraint-null} is the null geodesic. 
As for the conserved quantities along timelike geodesics \eqref{eq:ppn-constant-timelike}, there are two conserved quantities along null geodesics,
\eqb
\label{eq:ppn-constant-null}
 w \defeq -k_0 = \text{constant} \quad,\quad
 l_z \defeq x k_2 - y k_1 = \text{constant}
 \,.
\eqe
Physical meanings of $w$ and $l_z$ are respectively the energy and the angular momentum around z-axis of a photon propagating on the null geodesic, where $w$ has no dimension and $l_z$ has the length dimension. 
We do not expect the existence of a ``Carter-like'' constant, as discussed in Sect.\ref{subsubsec:ppn-timelike-hamiltonian}.

The \emph{parametrized post-Minkowskian (PPM)} expansion of $\Hk(x,k)$ is obtained by substituting the metric \eqref{eq:ppn-1PN1PM} into Eqs.\eqref{eq:ppn-hamiltonian-null},
\seqb
\label{eq:ppn-Hk-expansion}
\eqb
\label{eq:ppn-Hk-expansion-mainbody}
\begin{array}{rcll}
 \Hk(x,k)
 &=&
 \displaystyle
 \dfrac{1}{2}g^{00} w^2
 - g^{0j}\, k_j \, w
 + \dfrac{1}{2} g^{ij} k_i k_j
 &
\\[3mm]
 &=&
 \displaystyle
 - \dfrac{1}{2} w^2 + \dfrac{1}{2} \sum_{j=1}^3 k_j^2
 &\cdots\text{0PM : $O(1)$}
\\[4mm]
 &&
 - ( w^2 + B\, k\ls{r}^2 )\,\varepsilon(r)
 &\cdots\text{1PM : $O(\varepsilon)$}
\\[3mm]
 &&
 + O(\varepsilon^2)
 &\cdots\text{higher PM}
\end{array}
\eqe
where the order of terms is counted with only the PN parameter $\varepsilon(r)$ because the order of tangent 1-form is $O(k)=1$ for photons, and $\kr$ is given by
\eqb
\label{eq:ppn-kr}
 \kr \defeq \dfrac{1}{r} x^j k_j \,.
\eqe
\seqe
In Eq.\eqref{eq:ppn-Hk-expansion-mainbody}, ``$n$PM'' means the terms of $O(\varepsilon^{n})$. 
The 0PM terms express the null geodesic in the framework of the special relativity. 
The 1PM terms express the largest gravitational effect, which depends on the mass $m$ but not on the spin $a$. 
Note that Eq.\eqref{eq:app-Hk-expansion} in Appendix~\ref{app:ppn} shows the PPM expansion of $\Hk$ with retaining $\{ N\ls{t} , N\ls{s} , C_\perp \}$ in the metric \eqref{eq:ppn-all}.

\subsubsection{Geodesic equations}
\label{subsubsec:ppn-null-eq}

The PPM null geodesic equations are obtained by substituting Eqs.\eqref{eq:ppn-Hk-expansion} into Eq.\eqref{eq:ppn-hamiltoneq-null}. 
Here we summarize these equations with regarding the dynamical variables as the functions of the affine parameter $\sigma$, not of the coordinate time. 
\seqb
\label{eq:ppn-nullgeodesic}
\eqb
\label{eq:ppn-nullgeodesic-t}
\begin{array}{rcll}
 \od{t(\sigma)}{\sigma} &=& \pd{\Hk\bigl(\,x^\alpha(\sigma)\,,\,k_\alpha(\sigma)\,\bigr)}{(-w)} &
\\
 &=&
 w
 &
 \cdots\text{0PM}
\\[1mm]
 &&
 + 2 w \,\varepsilon(r)
 &
 \cdots\text{1PM}
\\[1mm]
 && 
 + O(\varepsilon^2)
 &
 \cdots\text{higher PM}
\end{array}
\eqe
\eqb
\label{eq:ppn-nullgeodesic-xj}
\begin{array}{rcll}
 \od{x^j(\sigma)}{\sigma} &=& \pd{\Hk\bigl(\,x^\alpha(\sigma)\,,\,k_\alpha(\sigma)\,\bigr)}{k_j(\sigma)} &
\\
 &=&
 k_j
 &
 \cdots\text{0PM}
\\[0mm]
 &&
 - 2 B k\ls{r} \dfrac{x^j}{r} \varepsilon(r)
 &
 \cdots\text{1PM}
\\[2mm]
 && 
 + O(\varepsilon^2)
 &
 \cdots\text{higher PM}
\end{array}
\eqe
\eqb
\label{eq:ppn-nullgeodesic-kj}
\begin{array}{rcll}
 \od{k_j(\sigma)}{\sigma} &=&
 -\pd{\Hk\bigl(\,x^\alpha(\sigma)\,,\,k_\alpha(\sigma)\,\bigr)}{x^j(\sigma)} &
\\
 &=& 0
 &
 \cdots\text{0PM}
\\[0mm]
 &&
 - \Bigr[\,
 w^2 \dfrac{x^j}{r}
 + B \Bigl( 3\dfrac{x^j}{r} k\ls{r}^2 - 2k_j k\ls{r} \Bigr)
 \,\Bigr]\,\dfrac{\varepsilon(r)}{r}
 &
 \cdots\text{1PM}
\\[3mm]
 && 
 + O(\varepsilon^2)
 &
 \cdots\text{higher PM}
 \,,
\end{array}
\eqe
\seqe
where $k_0 = -w$ is used. 
In Sect.\ref{subsec:obs-null}, from these geodesic equations, we will obtain analytic perturbative solutions of the 0PM and 1PM null geodesics which connect the star S0-2 and a distant observer representing us. 
As expected by the nullity of 0PM term of the acceleration \eqref{eq:ppn-nullgeodesic-kj}, the 0PM solution is of a constant velocity and corresponds to null geodesics on Minkowski metric.

\subsection{Coordinate system, initial condition of S0-2 and model parameters}
\label{subsec:ppn-setup}

\subsubsection{Coordinate system for observation}
\label{subsubsec:ppn-setup-XYZ}

Let us introduce the observer representing us so as to match with the actual observation process. 
In the reduction of observational values from observational raw data, the following effects are removed; the effect of earth's spin and revolution around the sun, and the effect of sun's peculiar motion with respect to the Local Standard of Rest (LSR) reference frame.
Therefore, we make our observer move with a velocity which is not removed in the above reduction process. 
The time scale of such observer's motion is expected to be of a time scale determined by the size of our galaxy $L\ls{gal}\sim 4\times 10^4$ pc, which gives $L\ls{gal}/c \sim 1.3\times 10^5$ years. 
It is thus appropriate to assume that the observer's relative velocity to \sgra is constant, because the time scale of S0-2 observations is of a few ten years which is very shorter than $1.3 \times 10^5$ years.

From the known approximated values of some parameters \eqref{eq:ppn-value}, the distance from our sun to \sgra is $R\ls{GC} \sim 2.4 \times 10^{14}$ km and the Schwarzschild radius of \sgra is $r\ls{sch} \sim 1.2 \times 10^7$ km. 
Then, the difference of time lapse between the sun and \sgra due to \sgra's gravity, which is estimated from the gravitational redshift, is $r\ls{sch}/R\ls{GC} \sim 5 \times 10^{-8}$. 
This means that, during 20 years observation from 2000 to 2020, a temporal difference of $20 \times r\ls{sch}/R\ls{GC} \sim 10^{-6}$ years arises between the sun and \sgra . 
Such small temporal uncertainty cannot be identified in the actual observation which needs about one day for obtaining one set of observational raw data. 
Further, when we estimate the magnitude of our observer's velocity $V\ls{obs}$ as an object bounded by \sgra's gravity, it becomes 
$V\ls{obs}/c \sim \varepsilon^{1/2}\bigr|_{r=R\ls{GC}} \sim (r\ls{sch}/R\ls{GC})^{1/2} \sim 10^{-4}$. 
The difference of time lapse between the sun and the coordinate time $t$ due to the velocity $V\ls{obs}$, which is estimate from the Lorentz factor, is $(V\ls{obs}/c)^2 \sim 10^{-8}$. 
This difference is also not identifiable in the present observational data. 
Thus we regard the coordinate time $t$ as the proper time of our observer.

\begin{figure}[t]
\begin{center}
\includegraphics[width=150mm]{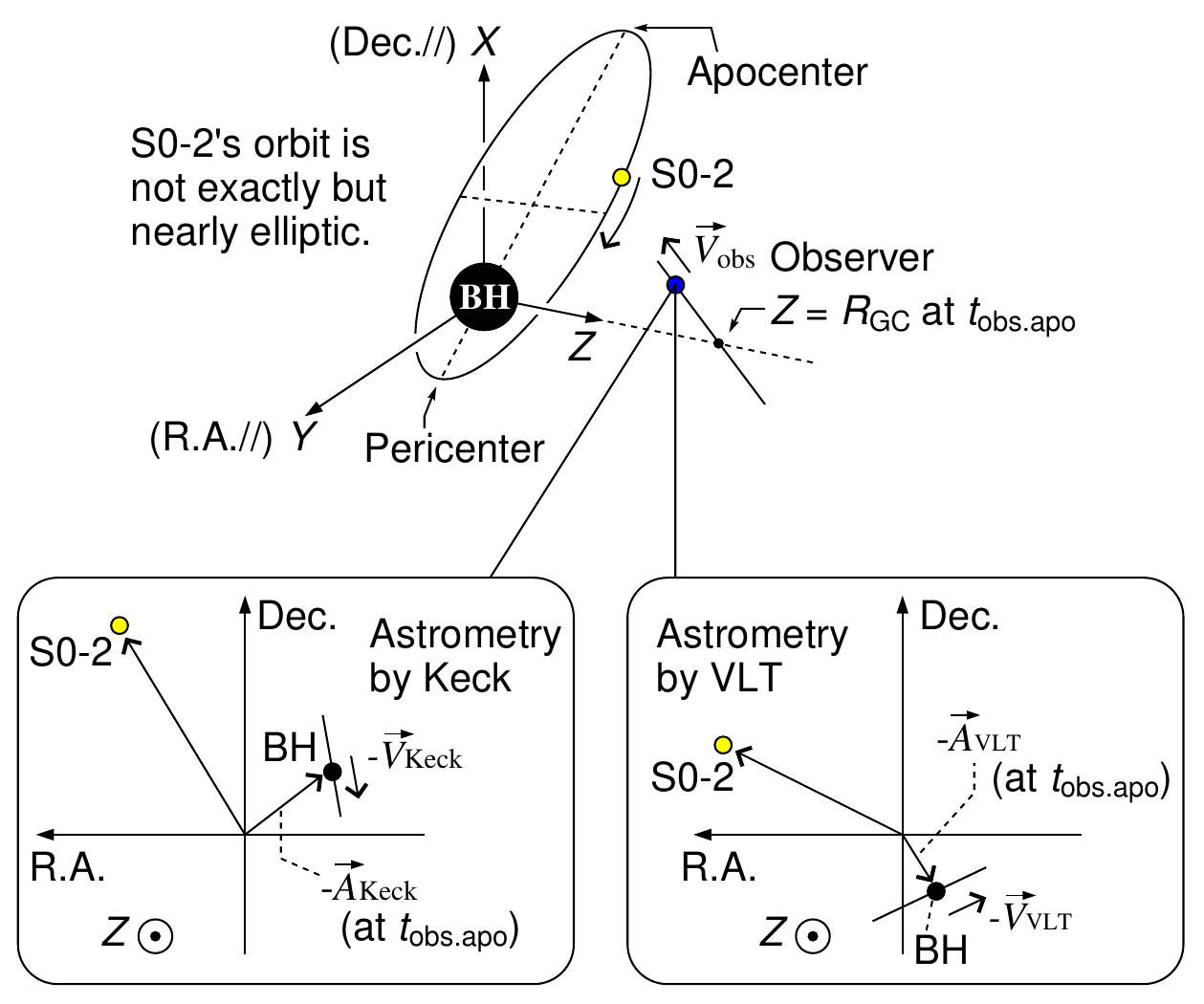}
\end{center}
\caption{
The observer representing us and the spatial coordinate system $(X,Y,Z)$ appropriate to the observation. 
``BH'' denotes \sgra. 
The observer's velocity $\vec{V}\ls{obs}$ is constant relative to \sgra. 
At the time $t\ls{obs.apo}$, the photon emitted by S0-2 at the apocenter reaches the observer (see Sect.\ref{subsubsec:ppn-setup-XYZ}). 
The origins of astrometry (observation of the stellar position on the sky plane) for Keck and VLT groups are assumed to be moving with constant velocity relative to \sgra (see Sect.\ref{subsubsec:ppn-setup-astrometry})
}
\label{fig:coordinates}
\end{figure}

The spatial coordinate system $(X,Y,Z)$ appropriate to the observation is introduced as shown in Fig.\ref{fig:coordinates}. 
We set $(X,Y,Z)$ be related with the Cartesian-like coordinates $(x,y,z)$ by a spatial rotation which will be explained in Sect.\ref{subsubsec:ppn-setup-spin}. 
The coordinate axes of $(X,Y,Z)$ are fixed by making use of the apocenter (the farthest point from \sgra) of the S0-2's orbit. 
\begin{itemize}
\item
The $Z$-axis points from \sgra to the spatial position of the observer where the photon emitted by S0-2 at the apocenter is received by the observer.
\item
The $Y$-axis points the same direction as the right ascension (R.A.), from the west to the east seen from the observer.
\item
The $X$-axis points the same direction as the declination (Dec.), from the south to the north seen from the observer.
\end{itemize}
In this observational coordinate system, we define the distance from the sun to \sgra, $R\ls{GC}$, as the $Z$ coordinate of the crossing event of the observer's orbit and $Z$-axis, at which the observer receives the photon emitted by S0-2 at the apocenter. 
Further, let $t\ls{obs.apo}$ denote the observation time of the photon emitted by S0-2 at the apocenter, which occurred already in 2010. 
Then, in the observational coordinates $(X,Y,Z)$, the spatial position of our observer $\vec{r}\ls{obs}(t)$ at a given observation time $t$ is given by
\eqb
\label{eq:ppn-observer}
 \vec{r}\ls{obs}(t) =
 (t-t\ls{obs.apo}) \vec{V}\ls{obs} + \vec{A}\ls{obs}
 \,,
\eqe
where $\vec{V}\ls{obs} = (V^X\ls{obs},V^Y\ls{obs},V^Z\ls{obs})$ is the constant velocity of the observer and $\vec{A}\ls{obs} = (0,0,R\ls{GC})$ in the observational coordinates $(X,Y,Z)$.

\subsubsection{Astrometric origin on the sky plan}
\label{subsubsec:ppn-setup-astrometry}

The visible 2D position of S0-2 on the sky plane (astrometric data) have been observed by European group (with VLT telescope) and American group (with mainly Keck telescope and partially Gemini telescope), while the redshift of photons coming from S0-2 (spectroscopic data) have been observed by those two groups and our Japanese group (with Subaru telescope). 
In Fig.\ref{fig:coordinates}, two sky planes for VLT and Keck groups are depicted. 
As explained below, the use of the astrometric data of S0-2 raises some additional parameters to be evaluated by fitting observational data and theoretical predictions.

The observations of S0-2 have to be performed by infrared astronomical observations, because stars at the center of our galaxy can be observable by infrared photons. 
Further, although \sgra itself is visible by radio waves radiated by very dilute plasma gases surrounding \sgra, the infrared photons from the gases are so faint that \sgra is not visible for infrared telescopes. 
This means that the origin of the 2D sky plane can not be set exactly at \sgra in infrared observations. 
In the actual astrometric observations, the origin of the sky plane is set at a position of, for example, an infrared flare event observed in the past in the vicinity of \sgra. 
The position of such past flare event is not exactly at \sgra and may be moving relative to \sgra. 
Therefore, we assume that the astrometric origin is moving relative to \sgra with a constant velocity. 
Further, because the setup of astrometric origins by VLT and Keck groups are not the same, the relative motion of the origin to \sgra should be introduced individually to the two astrometric data sets of VLT and Keck groups. 
Hence, the 2D displacement vector $\vec{O}_i(t)$ ($i =$ VLT, Keck) from \sgra to the astrometric origin on the sky plane at a given observation time $t$ is expressed as
\eqb
\label{eq:ppn-astrometry}
 \vec{O}\ls{VLT}(t) =
 (t-t\ls{obs.apo}) \vec{V}\ls{VLT} + \vec{A}\ls{VLT}
\quad,\quad
 \vec{O}\ls{Keck}(t) =
 (t-t\ls{obs.apo}) \vec{V}\ls{Keck} + \vec{A}\ls{Keck}
\eqe
where $\vec{V}_i = (V^X_i,V^Y_i)$ is the 2D constant velocity of the astrometric origin relative to \sgra, and $\vec{A}_i = (A^X_i,A^Y_i)$ is the 2D displacement of the astrometric origin from \sgra at $t\ls{obs.apo}$.

We consider that every astrometric data is the offset of the visible 2D position of S0-2 from the origin $\vec{O}_i(t)$ at observation time $t$. 
In fitting theoretical prediction with astrometric observational data, the theoretically calculated 2D position on the sky plane need to be corrected by $\vec{O}\ls{VLT}(t)$ for VLT's astrometric data and by $\vec{O}\ls{Keck}(t)$ for Keck's astrometric data. 
Then, the best-fitting values of the eight parameters of the two astrometric origins, 
$\{ V\ls{VLT}^X , V\ls{VLT}^Y , A\ls{VLT}^X , A\ls{VLT}^Y , V\ls{Keck}^X , V\ls{Keck}^Y , A\ls{Keck}^X , A\ls{Keck}^Y \}$, 
should be obtained at the same time with all the other parameters in our PPN modelling (see Sect.\ref{subsubsec:ppn-setup-parameters}).

Finally in this Sect.\ref{subsubsec:ppn-setup-astrometry}, let us make a comment on our treatment of the astrometric origin. 
When using only one astrometric data set of, for example, Keck group, we can require reasonably that the astrometric origin is moving with the observer, $\vec{V}\ls{Keck} = (V^X\ls{obs},V^Y\ls{obs})$. 
However, when using two astrometric data sets of both groups, we do not know how to fix the two velocities $\vec{V}_i$ ($i =$ VLT, Keck) in relation with $\vec{V}\ls{obs}$. 
In this paper, we leave the two velocities $\vec{V}_i$ as free parameters to be evaluated by fitting observational data and theoretical predictions.

\subsubsection{Black Hole's coordinate system}
\label{subsubsec:ppn-setup-spin}

\begin{figure}[t]
\begin{center}
\includegraphics[width=150mm]{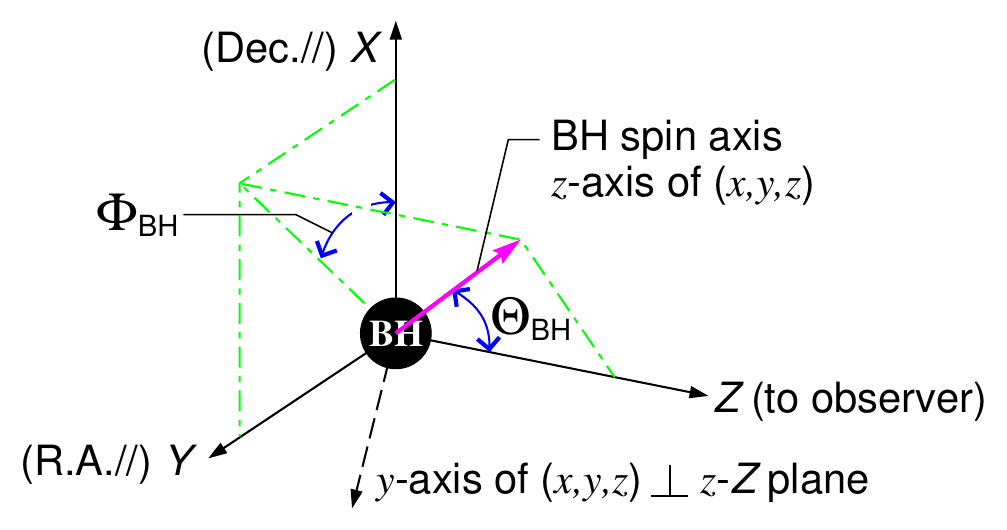}
\end{center}
\caption{
Directional angles $(\Theta\ls{BH},\Phi\ls{BH})$ of the spin axis of \sgra in the observational spatial coordinates $(X,Y,Z)$. 
``BH'' denotes \sgra. 
This spin axis is the $z$-axis of the Cartesian-like spatial coordinates $(x,y,z)$ which describe the metric \eqref{eq:ppn-1PN1PM}. 
These angles $(\Theta\ls{BH},\Phi\ls{BH})$ cannot be measured with the present observational uncertainties, and then we set $(\Theta\ls{BH},\Phi\ls{BH}) = (0,0)$ and $(X,Y,Z) = (x,y,z)$ in this paper. 
The spin effects are expected to be measurable by the near future telescope, for example Thirty-Meter-Telescope.
}
\label{fig:transform}
\end{figure}

As shown in Fig.\ref{fig:transform}, we determine the spatial rotation relating the observational spatial coordinates $(X,Y,Z)$ and the Cartesian-like spatial coordinates of the black hole $(x,y,z)$ by the following two conditions.
\begin{itemize}
\item
Let us express the direction of the spin axis of \sgra in the observational coordinates $(X,Y,Z)$ by the zenith and azimuth angles $(\Theta\ls{BH},\Phi\ls{BH})$ as shown in Fig.\ref{fig:transform}. 
This spin axis is the $z$-axis of the Cartesian-like coordinates which describe the metric \eqref{eq:ppn-1PN1PM}. 
\item
Let us fix the $y$-axis of the Cartesian-like coordinates $(x,y,z)$ in the observational coordinates $(X,Y,Z)$ so as to be parallel to the outer product $\vec{e}_Z \times \vec{e}_z$, where $\vec{e}_i$ denotes the unit spatial vector along $i$-axis and $i = z, Z$. 
Then, $x$-axis is automatically fixed as a right-handed system.
\end{itemize}
Under these conditions, the coordinate transformation is given by
\seqb
\label{eq:ppn-transformation-XYZxyz}
\eqb
 (X,Y,Z) = (x,y,z)\, {\cal T}[\Theta\ls{BH},\Phi\ls{BH}]
 \,,
\eqe
where ${\cal T}[\Theta\ls{BH},\Phi\ls{BH}]$ is a rotation matrix given by
\eqb
 {\cal T}[\Theta\ls{BH},\Phi\ls{BH}] =
 \left(
 \begin{array}{ccc}
 \cos\Theta\ls{BH} & 0 & -\sin\Theta\ls{BH}
 \\
 0 & 1 & 0
 \\
 \sin\Theta\ls{BH} & 0 &\cos\Theta\ls{BH}
 \end{array}
 \right)
 \left(
 \begin{array}{ccc}
 \cos\Phi\ls{BH} & \sin\Phi\ls{BH} & 0
 \\
 -\sin\Phi\ls{BH} & \cos\Phi\ls{BH} & 0
 \\
 0 & 0 & 1
 \end{array}
 \right) 
 \,.
\eqe
\seqe

It should be noted that the black hole spin effects of \sgra is not measurable by the present telescopes, and the spin magnitude $a$ and angles $(\Theta\ls{BH} , \Phi\ls{BH})$ cannot be evaluated. 
Therefore, in comparing observational data with theoretical predictions in this paper, we assume $(\Theta\ls{BH},\Phi\ls{BH}) = (0,0)$. 
Under this assumption, the coincidence of spatial coordinate systems $(X,Y,Z) \equiv (x,y,z)$ holds. 
When next generation telescopes, such as the Thirty-Meter-Telescope, starts scientific operations, the detection of the spin effect will be realized.

\subsubsection{Initial condition of S0-2's motion}
\label{subsubsec:ppn-setup-IC}

Given a initial time for calculating a stellar motion, the number of parameters for the initial condition of the stellar motion is six for the initial spatial position and the initial spatial velocity. 
We determine these six parameters for S0-2 as follows.

Let us note that the pericenter distance $r\ls{p}$ of S0-2 is about a thousand times the Schwarzschild radius $r\ls{sch}$ of \sgra, $r\ls{p} \sim 10^3 r\ls{sch}$, as indicated by Eq.\eqref{eq:ppn-pnparameter-peri}. 
Therefore the orbit of S0-2 is almost elliptic. 
Then, we set the initial condition at the apocenter observed in 2010. 
Because the radial component of S0-2's velocity at the apocenter vanishes, the number of parameters for the initial condition at the apocenter is reduced from six to five.

\begin{figure}[t]
\begin{center}
\includegraphics[width=150mm]{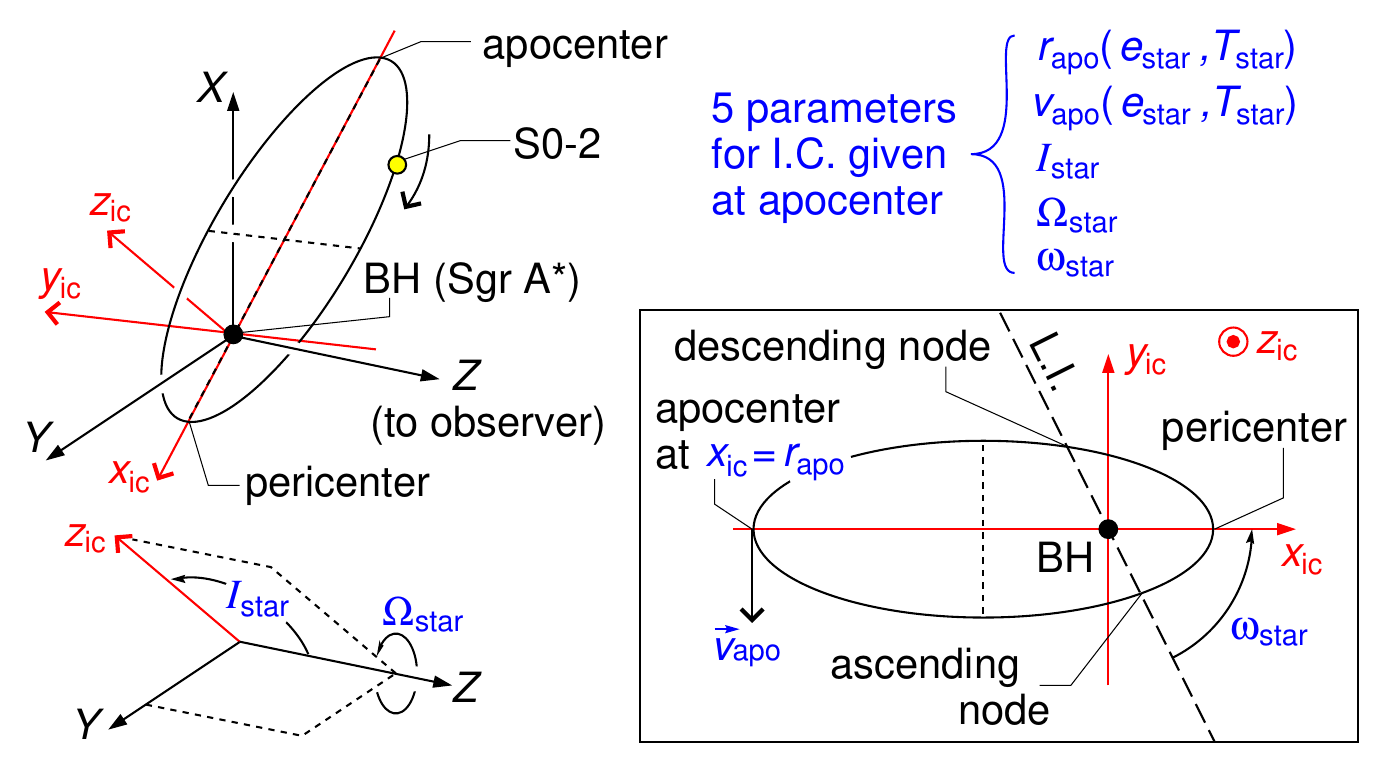}
\end{center}
\caption{
The initial condition of S0-2's motion at its apocenter observed in 2010. 
The orbit of S0-2 is almost elliptic, and a Keplerian elliptic motion can be imagined for given spatial position and velocity at the apocenter. 
This imaginary Keplerian orbit is on the $x\ls{ic}$-$y\ls{ic}$ plane. 
The apocenter distance $r\ls{apo}$ and speed $v\ls{apo}$ are given by the period $T\ls{star}$ and the eccentricity $e\ls{star}$ of the imaginary Keplerian orbit.  
Relation between two spatial coordinate systems $(x\ls{ic},y\ls{ic},z\ls{ic})$ and $(X,Y,Z)$ is described by three angles, $I\ls{star}$, $\Omega\ls{star}$ and $\omega\ls{star}$, where L.I. in the figure is the line of intersection of $X$-$Y$ plane and $x\ls{ic}$-$y\ls{ic}$ plane. 
The ascending node is the intersection point of L.I. and the stellar orbit, corresponding to the stellar velocity going away from the observer. 
}
\label{fig:ICapo}
\end{figure}

Given the initial spatial position and velocity at the apocenter, one can imagine a Keplerian elliptic motion which is determined by the given initial condition with assuming Newtonian gravity of \sgra. 
Further the difference between the imaginary Keplerian motion and the geodesic motion by Eqs.\eqref{eq:ppn-timelikegeodesic} is minimized, because the gravity of \sgra on the orbit of S0-2 becomes weakest at the apocenter. 
As shown in Fig.\ref{fig:ICapo}, with referring to the imaginary Keplerian motion, we introduce a spatial coordinate system $(x\ls{ic},y\ls{ic},z\ls{ic})$ relating with $(X,Y,Z)$ by a spatial rotation which will be given in Eqs.\eqref{eq:ppn-IC-XYZ}.
\begin{itemize}
\item
The $z\ls{ic}$-axis points the same direction as the spatial angular momentum of S0-2 at the apocenter observed in 2010. 
\item
The $x\ls{ic}$-axis points from the apocenter to the pericenter of the imaginary Keplerian elliptic orbit, 
\item
The $y\ls{ic}$-axis is automatically fixed as a right-handed system. 
\end{itemize}
In this coordinate system, the imaginary Keplerian orbit is on the $x\ls{ic}$-$y\ls{ic}$ plane.

The five parameters for the initial condition at the apocenter can be expressed by the five orbital parameters of the imaginary Keplerian motion; the orbital period $T\ls{star}$, the orbital eccentricity $e\ls{star}$, the inclination angle $I\ls{star}$, the ascending node angle $\Omega\ls{star}$, and the pericenter angle from the ascending node $\omega\ls{star}$. 
The definition of the angles $\{I\ls{star},\Omega\ls{star},\omega\ls{star}\}$ are shown in Fig.\ref{fig:ICapo}, and the other two parameters $\{T\ls{star},e\ls{star}\}$ are transformed to the apocenter distance $r\ls{apo}$ and speed $v\ls{apo}$ by the Keplerian formulas,
\eqb
\label{eq:ppn-IC-rv}
 r\ls{apo} =
 (1+e\ls{star})
 \left(\dfrac{T\ls{star} \sqrt{m}}{2\pi}
 \right)^{2/3}
\quad,\quad
 v\ls{apo} =
 \left(\dfrac{2\pi\, m}{T\ls{star}}
 \right)^{1/3}
 \sqrt{\frac{1-e\ls{star}}{1+e\ls{star}}
 }
 \,,
\eqe
where $m$ is the mass of black hole, $r\ls{apo}$ has the dimension of length, and $v\ls{apo}$ has no dimension. 
Note that, as will be shown in Sect.\ref{sec:fitting}, the difference between the Keplerian orbital period $T\ls{star}$ and the time interval between neighboring apocenters (or pericenters) in the framework of our PPN model is of a few days, while the duration of observational operation for obtaining one observational data is about one day. 
The Keplerian period $T\ls{star}$ is a good approximation as the observational orbital period.

From these five parameters $\{ r\ls{apo} , v\ls{apo} , I\ls{star}, \Omega\ls{star} , \omega\ls{star} \}$, the initial spatial position $(X\ls{apo},Y\ls{apo},Z\ls{apo})$ and velocity $(V^X\ls{apo},V^Y\ls{apo},V^Z\ls{apo})$ in the observational spatial coordinate system are calculated by
\seqb
\label{eq:ppn-IC-XYZ}
\eqb
\begin{split}
 (X\ls{apo},Y\ls{apo},Z\ls{apo})
 &=
 (-r\ls{apo},0,0) \, {\cal R}[I\ls{star},\Omega\ls{star},\omega\ls{star}]
 \\
 (V^X\ls{apo},V^Y\ls{apo},V^Z\ls{apo})
 &=
 (0,-v\ls{apo},0) \, {\cal R}[I\ls{star},\Omega\ls{star},\omega\ls{star}]
 \,,
\end{split}
\eqe
where ${\cal R}[I\ls{star},\Omega\ls{star},\omega\ls{star}]$ is a rotation matrix given by
\eqb
\begin{split}
 & {\cal R}[I\ls{star},\Omega\ls{star},\omega\ls{star}] =
 \\
 &
 \left(
 \begin{array}{ccc}
 \cos\omega\ls{star} & \sin\omega\ls{star} & 0
 \\
 -\sin\omega\ls{star} & \cos\omega\ls{star} & 0
 \\
 0 & 0 & 1
 \end{array}
 \right)
 \left(
 \begin{array}{ccc}
 1 & 0 & 0
 \\
 0 & \cos I\ls{star} & - \sin I\ls{star}
 \\
 0 & \sin I\ls{star} & \cos I\ls{star}
 \end{array}
 \right)
 \left(
 \begin{array}{ccc}
 \cos\Omega\ls{star} & \sin\Omega\ls{star} & 0
 \\
 -\sin\Omega\ls{star} & \cos\Omega\ls{star} & 0
 \\
 0 & 0 & 1
 \end{array}
 \right)
 \,.
\end{split}
\eqe
\seqe
The initial condition in the Cartesian-like coordinate system of black hole is given from the transformation \eqref{eq:ppn-transformation-XYZxyz},
\seqb
\label{eq:ppn-IC-xyz}
\eqab
 (x\ls{apo},y\ls{apo},z\ls{apo}) &=&
 (X\ls{apo},Y\ls{apo},Z\ls{apo}) \, {\cal T}[\Theta\ls{BH},\Phi\ls{BH}]^{-1}
 \\
 &=&
 (-r\ls{apo},0,0) \, {\cal R}[I\ls{star},\Omega\ls{star},\omega\ls{star}]
 \, {\cal T}[\Theta\ls{BH},\Phi\ls{BH}]^{-1}
 \nonumber
 \\
 (u^1\ls{apo},u^2\ls{apo},u^3\ls{apo}) &=&
 (V^X\ls{apo},V^Y\ls{apo},V^Z\ls{apo}) \, {\cal T}[\Theta\ls{BH},\Phi\ls{BH}]^{-1}
 \\
 &=&
 (0,-v\ls{apo},0) \, {\cal R}[I\ls{star},\Omega\ls{star},\omega\ls{star}]
 \, {\cal T}[\Theta\ls{BH},\Phi\ls{BH}]^{-1}
 \nonumber
 \,.
\eqae
The coordinate time $t\ls{star.apo}$ at which S0-2 passed the apocenter is given by
\eqb
\label{eq:ppn-IC-xyz-time}
 t\ls{star.apo} = t\ls{obs.apo} - \Delta t\ls{apo} \,,
\eqe
where $t\ls{obs.apo}$ is the time (in 2010) defined at Eq.\eqref{eq:ppn-observer}, and $\Delta t\ls{apo}$ is the propagation time of the photon from the apocenter of S0-2's orbit to the observer at $(t,X,Y,Z) = (t\ls{obs.apo},0,0,R\ls{GC})$. 
The concrete formula of $\Delta t\ls{apo}$ will be given from Eq.\eqref{eq:obs-propagationtime} in Sect.\ref{subsec:obs-null}. 
Further, with regarding the spatial velocity $u^j\ls{apo}$ as the spatial component of the initial four velocity, the temporal component is determined by the normalization condition $g_{\mu\nu}\us{apo}u^\mu\ls{apo} u^\nu\ls{apo} = -1$,
\eqb
\label{eq:ppn-IC-xyz-u0}
 u^0\ls{apo} =
 \dfrac{1}{g\us{apo}_{00}}
 \Bigl(\,
 - g\us{apo}_{0j}u\ls{apo}^j
 - \sqrt{(g\us{apo}_{0j}u\ls{apo}^j)^2
            - g\us{apo}_{00}\,\bigl( g\us{apo}_{jq}u\ls{apo}^j u\ls{apo}^q + 1 \bigr)}
 \,\Bigr)
 \,,
\eqe
where $g_{\mu\nu}\us{apo}$ is the metric tensor at the apocenter.
Note that the normalization condition is regarded as a second order algebraic equation of $u^0\ls{apo}$, whose two solutions are future pointing and past pointing. 
Eq.\eqref{eq:ppn-IC-xyz-u0} is the future pointing solution. 
Then, the future pointing initial condition for the 1-form $u_\mu\us{apo}$ is given by 
\eqb
 u_\mu\us{apo} = g_{\mu\nu}\us{apo} u^\nu\ls{apo}
 \,.
\eqe
\seqe
The timelike geodesic equations \eqref{eq:ppn-timelikegeodesic} are numerically integrated with the initial condition $x^\mu\ls{apo} = (t\ls{star.apo},x\ls{apo},y\ls{apo},z\ls{apo})$ and $u_\mu\us{apo}$.

\subsubsection{Parameters to be evaluated by observing S0-2}
\label{subsubsec:ppn-setup-parameters}

From the above, the model parameters in our PPN formulation are summarized in Table~\ref{table:parameters}.

\begin{table}[!h]
\caption{PPN model parameters to be evaluated by fitting predictions of the PPN model with observational data.}
\label{table:parameters}
\centering
\begin{tabular}{ccll}
\hline\hline
\multicolumn{4}{c}{parameters for BH/\sgra}
\\
\hline
Black hole mass &:&
$m$ &
see Eq.\eqref{eq:ppn-1PN1PM}
\\
Black hole spin &:&
$a$, $\Theta\ls{BH}$, $\Phi\ls{BH}$ &
see Eqs.\eqref{eq:ppn-1PN1PM}, \eqref{eq:ppn-transformation-XYZxyz}
\\
PPN parameters &:&
$A$, $B$, $C_z$ &
see Eq.\eqref{eq:ppn-1PN1PM}
\\
\hline\hline
\multicolumn{4}{c}{parameters for observer}
\\
\hline
Distance to \sgra &:&
$R\ls{GC}$ &
see Eq.\eqref{eq:ppn-observer}
\\
Observer's velocity &:&
$V^X\ls{obs}$, $V^Y\ls{obs}$, $V^Z\ls{obs}$ &
see Eq.\eqref{eq:ppn-observer}
\\
\hline\hline
\multicolumn{4}{c}{parameters for astrometric origin}
\\
\hline
Keck's astrometry &:&
$V^X\ls{Keck}$, $V^Y\ls{Keck}$, $A^X\ls{Keck}$, $A^Y\ls{Keck}$ &
see Eq.\eqref{eq:ppn-astrometry}
\\
VLT's astrometry &:&
$V^X\ls{VLT}$, $V^Y\ls{VLT}$, $A^X\ls{VLT}$, $A^Y\ls{VLT}$ &
see Eq.\eqref{eq:ppn-astrometry}
\\
\hline\hline
\multicolumn{4}{c}{parameters for S0-2's initial condition}
\\
\hline
Apocenter observation &:&
$t\ls{obs.apo}$ &
see Eqs.\eqref{eq:ppn-observer}, \eqref{eq:ppn-astrometry}, \eqref{eq:ppn-IC-xyz-time}
\\
Orbital period &:&
$T\ls{star}$ (Keplerian) &
see Eqs.\eqref{eq:ppn-IC-rv}, \eqref{eq:ppn-IC-xyz}
\\
Orbital eccentricity &:&
$e\ls{star}$ (Keplerian) &
see Eqs.\eqref{eq:ppn-IC-rv}, \eqref{eq:ppn-IC-xyz}
\\
Inclination angle &:&
$I\ls{star}$ (Keplerian) &
see Eq.\eqref{eq:ppn-IC-xyz}
\\
Ascending node angle &:&
$\Omega\ls{star}$ (Keplerian) &
see Eq.\eqref{eq:ppn-IC-xyz}
\\
Pericenter angle &:&
$\omega\ls{star}$ (Keplerian) &
see Eq.\eqref{eq:ppn-IC-xyz}
\\
\hline
\end{tabular}
\end{table}

One may think that the apocenter observation time $t\ls{obs.apo}$ is easily evaluated by observing continuously the motion of S0-2 near the apocenter passage. 
However such continuous observation is impossible in actual observations, and we can not necessarily obtain an observational data at the time $t\ls{obs.apo}$. 
Therefore, the apocenter observation time $t\ls{obs.apo}$ needs to be treated as a model parameter whose value should be estimated by fitting observational data and theoretical predictions.

Here let us note that, as will be estimated quantitatively in Sect.\ref{subsec:obs-PNPM}, the spin effects of \sgra are not measurable with the present observational uncertainties. 
The spin effects are expected to be measured by the next generation telescopes. 
Therefore, in fitting our PPN model with the present observational data, we fix the parameters for spin effects as follows.
\eqb
\label{eq:ppn-spin-undetectability}
 \text{Present undetectability of spin}: 
 \{a,\Theta\ls{BH},\Phi\ls{BH},C_z\}=\{0,0,0,0\} \,.
\eqe
Our PPN model under this assumption expresses a case that a star and photons move on geodesic orbits on a static spherically symmetric gravitational field.

\section{Observational quantities}
\label{sec:obs}

In this section, we derive the formulas of the following observational quantities as functions of the observational time $t$.
\begin{itemize}
\item
The offset of declination of S0-2 from \sgra, $\Delta\dec(t) = \ddec(t)$
\item
The offset of right ascension of S0-2 from \sgra, $\Delta\ra(t) = \dra(t)$
\item
The redshift of photons coming from S0-2, $\rs(t)$
\end{itemize}
These three observational quantities of S0-2 are being obtained by VLT, Keck and our Subaru groups. 
As explained in Sect.\ref{subsubsec:ppn-setup-astrometry}, in comparing the astrometric observables $(\ddec(t),\dra(t) )$ with the observational data, the offsets of the astrometric origins from \sgra given in Eq.\eqref{eq:ppn-astrometry} have to be added to the observational data, because the actual astrometric data express the offsets of S0-2's declination and right ascension from the astrometric origins.

The definitions of the three observational quantities are given by tetrad components of the null vector of photon detected by our observer \eqref{eq:ppn-observer}. 
In Sect.\ref{subsec:obs-null}, the analytic solutions of the PM null geodesic equations \eqref{eq:ppn-nullgeodesic} are obtained, and the propagation time $\Delta t$ of photons from S0-2 to the observer is also obtained. 
Then, in Sect.\ref{subsec:obs-radec} and \ref{subsec:obs-redshift}, the formulas of the three observational quantities are constructed by using the analytic PM solutions. 
In Sect.\ref{subsec:obs-PNPM}, the PN and PM orders which are detectable with the present telescopes are estimated.

\subsection{The null geodesic connecting S0-2 and observer, and the propagation time}
\label{subsec:obs-null}

In this section, we assume that the timelike geodesic equations \eqref{eq:ppn-timelikegeodesic} for the S0-2's motion have already been solved under the initial condition of S0-2's motion given in Sect.\ref{subsubsec:ppn-setup-IC}. 
Given the motion of S0-2, the null geodesics we need have to connect S0-2 and our observer. 
This means that we have to solve the ``boundary'' value problem of the null geodesic equations \eqref{eq:ppn-nullgeodesic}.

The affine parameter $\sigma$ of the null geodesics in Eqs.\eqref{eq:ppn-nullgeodesic} has the length dimension, so as to clarify the similarity with and difference from timelike geodesics. 
However, for the convenience of solving the boundary value problem, let us re-define the affine parameter to be non-dimensional by
\seqb
\label{eq:obs-redefinition}
\eqb
\label{eq:obs-sigmatilde}
 \sigma \to \sigmatilde \defeq \dfrac{\sigma}{\sigma_c} \,,
\eqe
where $\sigma_c$ is a constant of length dimension so as to satisfy
\eqb
 \sigmatilde =
 \begin{cases}
  0 & \text{at the emission of photon by S0-2}
  \\
  1 & \text{at the detection of photon by our observer}
 \end{cases}
 \,.
\eqe
The concrete value of $\sigma_c$ is not needed for calculating the observational quantities, as will be shown in Sect.\ref{subsec:obs-radec} and \ref{subsec:obs-redshift}. 
With adopting the new affine parameter $\sigmatilde$, the tangent vector of the null geodesic is also re-defined as
\eqb
\label{eq:obs-ktilde}
 k^\mu(\sigma) \to 
 \ktilde^\mu(\sigmatilde) \defeq
 \od{x^\mu(\sigmatilde)}{\sigmatilde} = \sigma_c k^\mu(\sigma)
 \,,
\eqe
\seqe
where $x^\mu(\sigmatilde)$ is the spacetime point on the null geodesic parametrized with $\sigmatilde$. 
This re-defined vector $\ktilde^\mu$ and the 1-form $\ktilde_\mu(\sigmatilde) = g_{\mu\nu}(\sigmatilde)\ktilde^\nu(\sigmatilde)$ have the dimension of lenght. 
The conserved quantities \eqref{eq:ppn-constant-null} are re-evaluated as
$\wtilde \defeq -\ktilde_0 = \sigma_c w$ (length dimension) and 
$\ltilde \defeq x\ktilde_2 - y\ktilde_1 = \sigma_c l_z$ (squared length dimension).

In order to solve Eqs.\eqref{eq:ppn-nullgeodesic}, we expand the dynamical variables $\{x^\mu(\sigmatilde),\ktilde_j(\sigmatilde)\}$ and $\wtilde$ as
\eqb
\label{eq:obs-expansion}
\begin{split}
 x^\mu(\sigmatilde) &=
 x^\mu\lzero(\sigmatilde) + x^\mu\lone(\sigmatilde) + O(r\ls{p}\varepsilon^2)
\\
 \ktilde_j(\sigmatilde) &=
 \ktilde_j\uzero(\sigmatilde) + \ktilde_j\uone(\sigmatilde) + O(\sigma_c \varepsilon^2)
\\
 \wtilde &= \wtilde\uzero + \wtilde\uone + O(\sigma_c \varepsilon^2)
 \,, 
\end{split}
\eqe
where $r\ls{p}$ is the spatial distance between \sgra and the photon at the point $x^\mu(\sigmatilde)$, and the suffix $(n)$ denotes $n$PM terms of orders $x^\mu_{(n)}(\sigmatilde) \sim O(r\ls{p}\varepsilon^n)$ and $\ktilde_\mu^{(n)} \sim O(\sigma_c\varepsilon^n)$. 
Note that, even if some PM terms $\wtilde^{(n)}$ of $\wtilde$ may by functions of $\sigmatilde$, the summation of those terms produces the constant $\wtilde$. 
By substituting the expansion \eqref{eq:obs-expansion} into Eqs.\eqref{eq:ppn-nullgeodesic}, we obtain analytic 0PM and 1PM solutions.

\subsubsection{0PM solution}
\label{subsubsec:obs-null-0PM}

The 0PM order terms of Eqs.\eqref{eq:ppn-nullgeodesic} are
\seqb
\label{eq:obs-0PM-equation}
\eqb
\label{eq:obs-0PM-equation-differential}
 \od{t\lzero(\sigmatilde)}{\sigmatilde} = \wtilde\uzero
 \quad,\quad
 \od{x\lzero^j(\sigmatilde)}{\sigmatilde} = \ktilde\uzero_j
 \quad,\quad
 \od{\ktilde\uzero_j}{\sigmatilde} = 0
 \,,
\eqe
where the null condition \eqref{eq:ppn-cosntraint-null} at 0PM order gives
\eqb
\label{eq:obs-0PM-equation-wtilde0}
 \wtilde\uzero = - \ktilde\uzero_0 =
 \sqrt{\sum_{j=1}^3 \bigl(\ktilde\uzero_j\bigr)^2}
 \,.
\eqe
\seqe
The appropriate boundary condition of the 0PM solution parametrized by $\sigmatilde$ consists of the following seven requirements,
\eqb
\label{eq:obs-0PM-BC}
 t\lzero(0) = t\ls{emi} \quad,\quad
 x\lzero^j(0) = x\ls{emi}^j \quad.\quad
 x\lzero^j(1) = x\ls{obs}^j \,,
\eqe
where $t\ls{emi}$ is the time coordinate at which S0-2 emits the photon, 
$x\ls{emi}^j \defeq x\ls{star}^j(t\ls{emi})$ 
is the spatial coordinate of S0-2 given by the solution of the timelike geodesic equations~\eqref{eq:ppn-timelikegeodesic}, and 
$x\ls{obs}^j = (x\ls{obs},y\ls{obs},z\ls{obs}) \defeq \vec{r}\ls{obs}(t\ls{obs(0)}) {\cal T}[\Theta\ls{BH},\Phi\ls{BH}]^{-1}$ is the spatial coordinate of our observer~\eqref{eq:ppn-observer} at the 0PM observation time $t\ls{obs(0)}$. 
Here note that the time $t\ls{obs(0)}$ needs to be determined by using the 0PM solution $t\lzero(\sigmatilde)$ as $t\ls{obs(0)} = t\lzero(1)$.

\begin{figure}[t]
\begin{center}
\includegraphics[width=150mm]{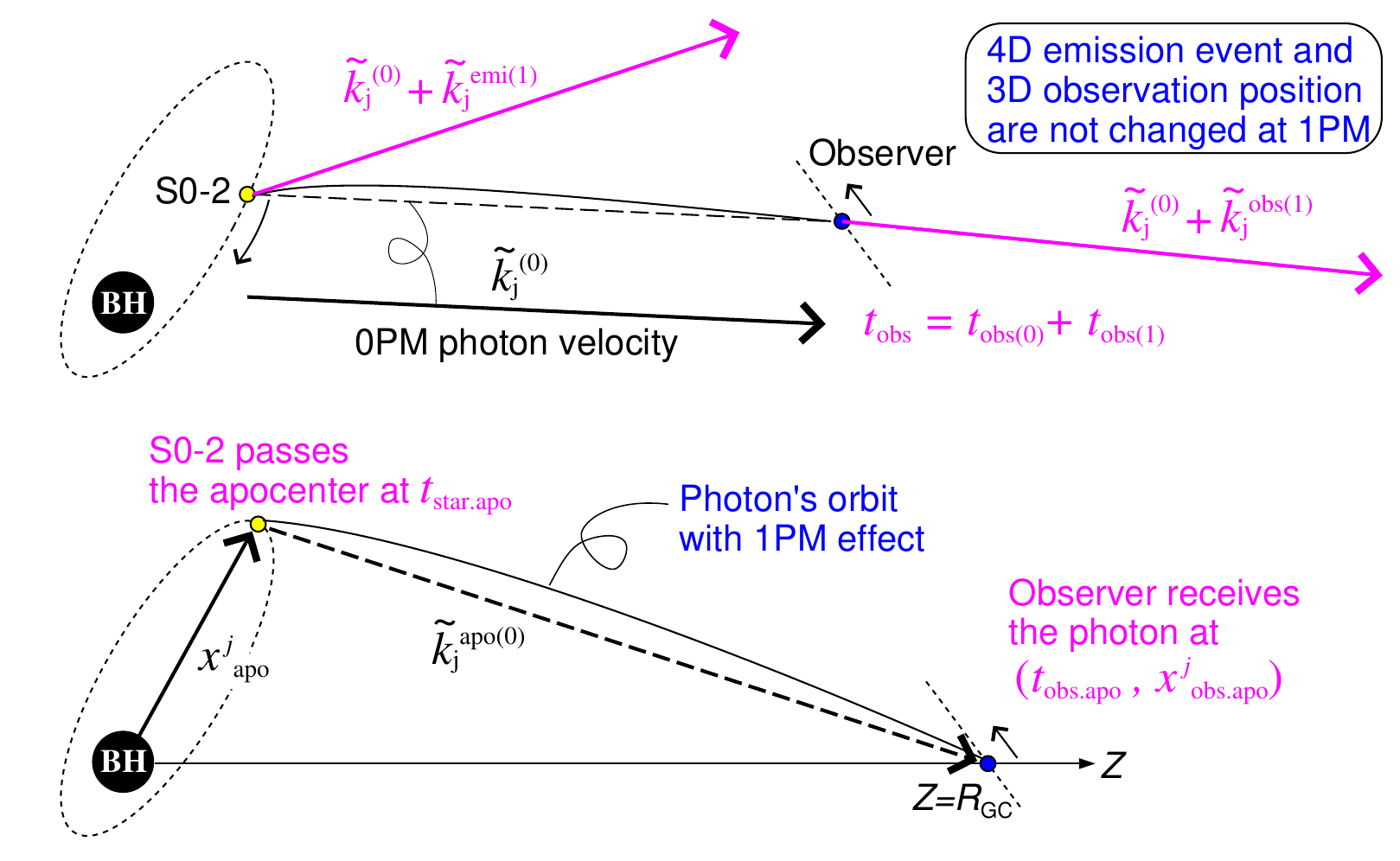}
\end{center}
\caption{
1PM correction of null geodesics. 
The emission event $(t\ls{emi}, x\ls{emi}^j)$ and the observation position $x\ls{obs}^j$ of photons are fixed, while the observation time $t\ls{obs} = t\ls{obs(0)} + t\ls{obs(1)}$ and the null 1-form $\ktilde_\mu = \ktilde\uzero_\mu + \ktilde\uone_\mu$ are corrected from 0PM case to 1PM case. 
The spatial part of 0PM 1-form $\ktilde\uzero_j$ is the ``positional vector'' connecting from $x\ls{emi}^j$ to $x\ls{obs}^j$. 
As defined in Sect.\ref{subsubsec:ppn-setup-XYZ}, the $Z$-axis of coordinates $(X,Y,Z)$ passes the observation position $x\ls{obs.apo}^j$ of the photon emitted at the apocenter passage event of S0-2. 
}
\label{fig:1PMphoton}
\end{figure}

These conditions and Eqs.\eqref{eq:obs-0PM-equation-differential} denote that the 0PM spatial vector $\ktilde\uzero\,^j = \ktilde\uzero_j$ is a ``positional vector'' connecting from $x\ls{emi}^j$ to $x\ls{obs}^j$ as shown in the upper panel of Fig.\ref{fig:1PMphoton}, 
\eqb
\label{eq:obs-ktilde0-primitive}
 \ktilde\uzero_j =
 (\wtilde\uzero + t\ls{emi} - t\ls{obs.apo}) V\ls{obs}^j + x\ls{obs.apo}^j - x\ls{emi}^j
 \,,
\eqe
where $V\ls{obs}^j = (V\ls{obs}^x,V\ls{obs}^y,V\ls{obs}^z) = (V\ls{obs}^X,V\ls{obs}^Y,V\ls{obs}^Z) {\cal T}[\Theta\ls{BH},\Phi\ls{BH}]^{-1}$ is the constant velocity of our observer \eqref{eq:ppn-observer}, and $x\ls{obs.apo}^j = (0,0,R\ls{GC}) {\cal T}[\Theta\ls{BH},\Phi\ls{BH}]^{-1}$ is the spatial position of our observer at $t\ls{obs.apo}$. 
Substituting Eq.\eqref{eq:obs-0PM-equation-wtilde0} into the right-hand side of Eq.\eqref{eq:obs-ktilde0-primitive}, $\ktilde\uzero_j$ is obtained. 

Eqs.\eqref{eq:obs-ktilde0-primitive} and \eqref{eq:obs-0PM-equation-wtilde0} become a quadratic equation of $\ktilde\uzero_j$. 
From the two solutions of it, we choose the solution which reduces to 
$\ktilde\uzero_j = x\ls{obs.apo}^j - x\ls{apo}^j$ 
at $t\ls{emi} = t\ls{star.apo(0)}$, 
where $x\ls{apo}^j$ is the spatial position of the apocenter of S0-2's orbit and $t\ls{star.apo(0)}$ is the apocenter passage time of S0-2 evaluated with 0PM photon propagation.\footnote{
This time is given by $t\ls{star.apo(0)} = t\ls{obs.apo} - \wtilde\uzero\ls{apo}$, where $\wtilde\uzero\ls{apo}$ is the 0PM conserved quantity of photon emitted from S0-2 at the apocenter. 
This $\wtilde\uzero\ls{apo}$ gives the 0PM propagation time of photon from the apocenter to our observer, and satisfies 
$\wtilde\ls{apo}\us{(0)\,2} = \sum_j (x\ls{obs.apo}^j - x\ls{apo}^j)^2$. 
} 
Consequently the analytic 0PM solutions of Eqs.\eqref{eq:obs-0PM-equation} are
\seqb
\label{eq:obs-0PM-sol}
\eqb
\begin{split}
 t\lzero(\sigmatilde) &= \wtilde\uzero \sigmatilde + t\ls{emi}
 \\
 x\lzero^j(\sigmatilde) &= \ktilde\uzero_j \sigmatilde + x\ls{emi}^j
 \\
 \ktilde\uzero_j &=
 {\mathcal D}\vec{x}
 + \dfrac{V\ls{obs}^j}{1-V\ls{obs}^2}
    \,\Bigl[\,
    - {\mathcal D}t
    + \vec{V}\ls{obs}\cdot{\mathcal D}\vec{x}
    + \sqrt{\heartsuit}
    \,\Bigr]
 \,,
\end{split}
\eqe
and the 0PM observation time becomes
\eqb
\label{eq:obs-0PM-sol-tobs0}
 t\ls{obs(0)} = t\lzero(1) =\wtilde\uzero + t\ls{emi}
 \,,
\eqe
where the spatial vectors $\vec{V}\ls{obs}$ and ${\mathcal D}\vec{x}$ in the above solution are the collection of spatial components in the Cartesian-like coordinates and
\eqb
\begin{split}
 {\mathcal D}t
 &\defeq
 t\ls{obs.apo} - t\ls{emi}
\\
 {\mathcal D}\vec{x}
 &\defeq
 \big(\,
 x\ls{obs.apo}-x\ls{emi} \,,\,
 y\ls{obs.apo}-y\ls{emi} \,,\,
 z\ls{obs.apo}-z\ls{emi}
 \bigr)
\\
 \heartsuit
 &\defeq
 \bigl(\,
  {\mathcal D}\vec{x} - {\mathcal D}t\,\vec{V}\ls{obs}
  + \vec{V}\ls{obs}\times{\mathcal D}\vec{x}
 \,\bigr)
 \cdot
 \bigl(\,
  {\mathcal D}\vec{x} - {\mathcal D}t\,\vec{V}\ls{obs}
  - \vec{V}\ls{obs}\times{\mathcal D}\vec{x}
 \,\bigr)
 \,,
\end{split}
\eqe
\seqe
where 
$\vec{a}\cdot\vec{b} \defeq a^1 b^1 + a^2 b^2 + a^3 b^3$ 
and
$\vec{a}\times\vec{b} \defeq
 \big(\,
 a^2 b^3 - a^3 b^2 \,,\,  a^3  b^1 - a^1 b^3 \,,\,  a^1 b^2- a^2 b^1
 \,\bigr)$
for any spatial vectors $\vec{a} = (a^1,a^2,a^3)$ and $\vec{b} = (b^1,b^2,b^3)$.

\subsubsection{1PM solution}
\label{subsubsec:obs-null-1PM}

The 1PM order terms of Eqs.\eqref{eq:ppn-nullgeodesic} are
\seqb
\label{eq:obs-1PM-equation}
\eqb
\label{eq:obs-1PM-equation-differential}
\begin{split}
 \od{t\lone(\sigmatilde)}{\sigmatilde}
 &=
 \wtilde\uone + 2 \wtilde\uzero \varepsilon\lzero
\\
 \od{x\lone^j(\sigmatilde)}{\sigmatilde}
 &=
 \ktilde\uone_j
 -2 B \ktilde\ls{r}\uzero \dfrac{x\lzero^j}{r\lzero} \varepsilon\lzero
\\
 \od{\ktilde\uone_j(\sigmatilde)}{\sigmatilde}
 &=
 -
 \Bigl[\,
  \bigl(\wtilde\uzero\bigr)^2\, \dfrac{x\lzero^j}{r\lzero}
  + B \Bigl(\, 3 \bigl(\ktilde\ls{r}\uzero\bigr)^2\, \dfrac{x\lzero^j}{r\lzero}
                 -2 \ktilde\ls{r}\uzero\, \ktilde\uzero_j \,\Bigr)
 \,\Bigr]\,\dfrac{\varepsilon\lzero}{r\lzero}
 \,,
\end{split}
\eqe
where 
$r\lzero(\sigmatilde) = \sqrt{x\lzero^2+y\lzero^2+z\lzero^2}$, 
$\varepsilon\lzero(\sigmatilde) = m/r\lzero$, 
$\ktilde\uzero\ls{r}(\sigmatilde) = r\lzero^{-1} x\lzero^j \ktilde\uzero_j$, 
and the null condition \eqref{eq:ppn-cosntraint-null} at 1PM order gives
\eqb
\label{eq:obs-1PM-equation-wtilde0}
 \wtilde\uone(\sigmatilde) =
 -\ktilde\uone_0(\sigmatilde) =
 \dfrac{1}{2 \wtilde\uzero}\Bigl(\,
   g\lone^{\mu\nu} \ktilde\uzero_\mu \ktilde\uzero_\nu
 + 2 \sum_{j=1}^3 \ktilde\uzero_j \ktilde\uone_j
 \,\Bigr)
 \,,
\eqe
where $g\lone^{\mu\nu}(\sigmatilde)$ is the terms of $O(\varepsilon)$ in the inverse metric \eqref{eq:ppn-1PN1PM-inverse}, 
\eqb
 g\lone^{00}(\sigmatilde)
 = -2 \varepsilon\lzero
 \quad,\quad
 g\lone^{0j} = 0
 \quad,\quad
 g\lone^{ij}(\sigmatilde)
 = -2 B \dfrac{x\lzero^i x\lzero^j}{r\lzero^2} \varepsilon\lzero
 \,.
\eqe
\seqe
The appropriate boundary condition of the 1PM solution consists of the followings,
\eqb
\label{eq:obs-1PM-BC}
 t\lone(0) = 0 \quad,\quad
 x\lone^j(0) = 0 \quad.\quad
 x\lone^j(1) = 0
 \,.
\eqe
This boundary condition denotes that, as shown in Fig.\ref{fig:1PMphoton}, the emission event of photon by S0-2 $x\ls{emi}^\mu$ and the spatial observation position of  our observer $x\ls{obs}^j$ are the same with 0PM case, while the observation time is corrected as $t\ls{obs} = t\ls{obs(0)} + t\ls{obs(1)}$, where $t\ls{obs(1)} = t\lone(1)$.

Under the above conditions, the 1PM equations \eqref{eq:obs-1PM-equation}, with substituting the 0PM analytic solutions \eqref{eq:obs-0PM-sol}, can be integrated analytically. 
In Appendix~\ref{app:1PM}, a few notes on this integration is summarized. 
Further, let us emphasize that the necessary information for calculating the observational quantities are the observation time $t\ls{obs(1)}$, and the null 1-forms at the emission $\ktilde\us{emi(1)}_\mu = \ktilde\uone_\mu(0)$ and at the observation $\ktilde\us{obs(1)}_\mu = \ktilde\uone_\mu(1)$. 
These necessary 1PM quantities, which are obtained by integrating Eqs.\eqref{eq:obs-1PM-equation-differential}, are as follows.
\seqb
\label{eq:obs-1PM-sol}
\eqab
\label{eq:obs-1PM-sol-kemi1}
\nonumber
 \ktilde\us{emi(1)}_j &=&
 - 2 B m \,\Bigl(\, \dfrac{x\ls{obs}^j}{r\ls{obs}} - \dfrac{x\ls{emi}^j}{r\ls{emi}} \,\Bigr)
\\
 &&
 - m\, \ktilde\uzero_j \,
   \Biggl[\,
     \dfrac{1-B}{\wtilde\uzero}
     \ln\Bigl|\,
     \dfrac{\wtilde\uzero\,^2 + \wtilde\uzero r\ls{obs} + b\uzero}
     {\wtilde\uzero r\ls{emi} + b\uzero}
     \,\Bigr|
\\
\nonumber
 &&\phantom{- m\, \ktilde\uzero_j \,\Bigl[\,}
   - \dfrac{B}{\wtilde\uzero\,^2}
      \Bigl(\,
      \dfrac{\wtilde\uzero\,^2 + b\uzero}{r\ls{obs}}
      - \dfrac{b\uzero}{r\ls{emi}}
      + \dfrac{b\uzero\,^2}{r\ls{emi}^3}
      \,\Bigr)
   - \dfrac{1}{r\ls{emi}}
   \,\Biggr]
\\
\nonumber
 &&
 - m\, q\uzero_j \,
   \Biggl[\,
     \dfrac{1+B}{r\ls{emi}^2-(b\uzero/\wtilde\uzero)^2}
     \Bigl(\, r\ls{obs}-r\ls{emi} + \dfrac{b\uzero}{r\ls{emi}} \,\Bigr)
   - B\,\Bigl(\,
   \dfrac{1}{r\ls{obs}}-\dfrac{1}{r\ls{emi}} + \dfrac{b\uzero}{r\ls{emi}^3} \,\Bigr)
   \,\Biggr]
\eqae
\eqab
\label{eq:obs-1PM-sol-kobs1}
 \ktilde\us{obs(1)}_j &=& \ktilde\us{emi(1)}
\\
\nonumber
 &&
 + m\, \ktilde\uzero_j \,
   \Biggl[\,
     (1+B)\,\Bigl(\,\dfrac{1}{r\ls{obs}} - \dfrac{1}{r\ls{emi}} \,\Bigr)
   - B \Bigl(\, r\ls{emi}^2 - \Bigl(\dfrac{b\uzero}{\wtilde\uzero}\Bigr)^2\,\Bigl)
      \,\Bigl(\,\dfrac{1}{r\ls{obs}^3} - \dfrac{1}{r\ls{emi}^3} \,\Bigr)
   \,\Biggr]
\\
\nonumber
 &&
 + m\, q\uzero_j \,
   \Biggl[\,
   - \dfrac{1+B}{r\ls{emi}^2-(b\uzero/\wtilde\uzero)^2}
     \Bigl(\,\dfrac{\wtilde\uzero\,^2 + b\uzero}{r\ls{obs}} - \dfrac{b\uzero}{r\ls{emi}} \,\Bigr)
\\
\nonumber
 &&\phantom{+ m\, q\uzero_j \,\Biggl[\,}
   + B 
     \Bigl(\,
     \dfrac{\wtilde\uzero\,^2 + b\uzero}{r\ls{obs}^3} - \dfrac{b\uzero}{r\ls{emi}^3}
     \,\Bigr)
   \,\Biggr]
\eqae
\eqb
\label{eq:obs-1PM-sol-tobs1}
 t\ls{obs(1)} =
 - B \dfrac{m}{\wtilde\uzero}
   \Bigl(\,\dfrac{\wtilde\uzero\,^2 + b\uzero}{r\ls{obs}} - \dfrac{b\uzero}{r\ls{emi}} \,\Bigr)
 + (1+B)\,m\,
   \ln\Bigl|\,
       \dfrac{\wtilde\uzero\,^2 + \wtilde\uzero r\ls{obs} + b\uzero}
       {\wtilde\uzero r\ls{emi} + b\uzero}
       \,\Bigr|
  \,,
\eqe
where $r\ls{emi} = r\lzero(0)$, $r\ls{obs} = r\lzero(1)$, and
\eqb
\label{eq:obs-1PM-sol-b0q0}
 b\uzero \defeq x\ls{emi}^j \ktilde\uzero_j
 \quad,\quad
 q\uzero_j \defeq
 x\ls{emi}^j - \dfrac{b\uzero}{\wtilde\uzero\,^2} \ktilde\uzero_j
 \,.
\eqe
\seqe
This $q\uzero_j$ is the perpendicular part of $x\ls{emi}^j$ to $\ktilde\uzero_j$ as implied by an identity, $\sum_j q\uzero_j \ktilde\uzero_j \equiv 0$. 
The propagation time of photon $\Delta t$ up to 1PM order is given by
\eqb
\label{eq:obs-propagationtime}
 \Delta t \simeq t\ls{obs(0)} + t\ls{obs(1)} - t\ls{emi}
 \,,
\eqe
and the term $\Delta t\ls{apo}$ in Eq.\eqref{eq:ppn-IC-xyz-time} is given by evaluating this $\Delta t$ at the apocenter passage of S0-2. 
Concerning these 1PM solutions, let us make two notes:
\begin{itemize}
\item
In the 1PM solutions \eqref{eq:obs-1PM-sol}, the PPN parameter $B$ appears explicitly, while the other PPN parameters $A$ and $C_z$ do not. 
Because the 1PM null geodesic equations \eqref{eq:obs-1PM-equation} includes $B$ but not $A$ and $C_z$. 
\item
The PPN parameters $A$ and $C_z$ affect the 1PM solutions implicitly through the emission event of photon $x\ls{emi}^\mu$ which is determined by the S0-2's motion. 
Because the S0-2's motion is determined by  the timelike geodesic equations \eqref{eq:ppn-timelikegeodesic} which depend on $A$ and $B$ at 1PN order and on $C_z$ at 1.5PN order. 
\end{itemize}

\subsection{Astrometric observables: Right Ascension and Declination}
\label{subsec:obs-radec}

Let us proceed to define the astrometric observables; the offsets of declination and right ascension of S0-2 from \sgra, $\Delta\dec = \ddec$ and $\Delta\ra = \dra$. 
They are defined with the tetrad components of the observed photon's four velocity vector $k\ls{obs}^{(I)} = \sigma_c^{-1} \ktilde\ls{obs}^{(I)}$ ($I = t,X,Y,Z$) aligned with the axes of observational coordinates $(t,X,Y,Z)$. 
The observables are defined as
\eqb
\label{eq:obs-astrometry-def}
 \ddec \defeq \arctan\dfrac{k^{(X)}\ls{obs}}{k^{(Z)}\ls{obs}}
 = \arctan\dfrac{\ktilde^{(X)}\ls{obs}}{\ktilde^{(Z)}\ls{obs}}
\quad,\quad
 \dra \defeq \arctan\dfrac{k^{(Y)}\ls{obs}}{k^{(Z)}\ls{obs}}
 = \arctan\dfrac{\ktilde^{(Y)}\ls{obs}}{\ktilde^{(Z)}\ls{obs}}
 \,.
\eqe
These are not affected by the value of $\sigma_c$. 
Here note that, because the gravity of \sgra can be ignored at our observer as discussed in Sect.\ref{subsubsec:ppn-setup-XYZ}, the tetrad components $\ktilde\ls{obs}^{(I)}$ are regarded as the coordinate components $\ktilde\ls{obs}^I = \eta^{IJ} \ktilde\us{obs}_J$ in the observational coordinates, where $\eta^{IJ} = {\rm diag}(-1,1,1,1)$. 
Further, by the coordinate transformation \eqref{eq:ppn-transformation-XYZxyz}, the spatial components of this null vector is calculated as
\eqb
\label{eq:obs-kXYZ}
 \bigl(\,
 \ktilde\ls{obs}^X \,,\,
 \ktilde\ls{obs}^Y \,,\,
 \ktilde\ls{obs}^Z 
 \,\bigr)
 =
 \bigl(\,
 \ktilde\us{obs}_X \,,\,
 \ktilde\us{obs}_Y \,,\,
 \ktilde\us{obs}_Z 
 \,\bigr)
 =
 \bigl(\,
 \ktilde\us{obs}_x \,,\,
 \ktilde\us{obs}_y \,,\,
 \ktilde\us{obs}_z 
 \,\bigr)
 {\cal T}[\Theta\ls{BH},\Phi\ls{BH}]
 \,,
\eqe
where the null 1-form at our observer $\ktilde\us{obs}_j = \ktilde\uzero_j + \ktilde\us{obs(1)}_j + O(\sigma_c \varepsilon^2)$ in $(x,y,z)$ coordinates are already given in Eqs.\eqref{eq:obs-0PM-sol} and \eqref{eq:obs-1PM-sol}.

Under the present undetectablity of BH's spin \eqref{eq:ppn-spin-undetectability}, we set $\{\Theta\ls{BH},\Phi\ls{BH}\} =\{0,0\}$ and then the PM expansion of the astrometric observables are obtained,
\seqb
\label{eq:obs-astrometry}
\eqb
 \ddec = \ddec\us{(0)} + \ddec\us{(1)} + O(\varepsilon^2)
\quad,\quad
 \dra = \dra\us{(0)} + \dra\us{(1)} + O(\varepsilon^2)
 \,,
\eqe
where the terms of $O(1)$ are
\eqb
 \ddec\us{(0)} = \dfrac{\ktilde\uzero_x}{\ktilde\uzero_z}
\quad,\quad
 \dra\us{(0)} = \dfrac{\ktilde\uzero_y}{\ktilde\uzero_z}
 \,,
\eqe
and the terms of $O(\varepsilon)$ are
\eqb
 \ddec\us{(1)} =
 \dfrac{\ktilde\us{obs(1)}_x}{\ktilde\uzero_z}
 - \dfrac{\ktilde\uzero_x}{\ktilde\uzero_z} \dfrac{\ktilde\us{obs(1)}_z}{\ktilde\uzero_z}
\quad,\quad
 \dra\us{(1)} =
 \dfrac{\ktilde\us{obs(1)}_y}{\ktilde\uzero_z}
 - \dfrac{\ktilde\uzero_y}{\ktilde\uzero_z} \dfrac{\ktilde\us{obs(1)}_z}{\ktilde\uzero_z}
 \,.
\eqe
\seqe
It should be emphasized that the unit of $\ddec$ and $\dra$ in Eq.\eqref{eq:obs-astrometry} is radian, while the unit of astrometric observational values is usually arcsec.

\subsection{Spectroscopic observable: Redshift}
\label{subsec:obs-redshift}

The redshift of photons coming from S0-2 to our observer, $\rs(t)$, is defined from the frequency at the emission by S0-2, $\nu\ls{emi}$, and that at the observation by our observer, $\nu\ls{obs}$,
\eqb
\label{eq:obs-redshift-def}
 \rs(t\ls{obs}) \defeq \dfrac{\nu\ls{emi}(t\ls{emi})}{\nu\ls{obs}(t\ls{obs})} - 1
 \,,
\eqe
where $t\ls{emi}$ is the emission time of the photon which is determined by the observation time $t\ls{obs}$, and up to the 1PM order of photon's propagation $t\ls{obs} = t\ls{obs(0)} + t\ls{obs(1)}$, Eq.\eqref{eq:obs-0PM-sol-tobs0} gives $t\ls{emi} = t\ls{obs} - \wtilde\uzero - t\ls{obs(1)}$. 
The frequencies in the definition \eqref{eq:obs-redshift-def} are given by
\eqb
\label{eq:obs-frequency-def}
\begin{split}
 \nu\ls{emi}(t\ls{emi}) &\defeq
 - u\ls{star}^\mu k\us{emi}_\mu =
 - \sigma_c^{-1} u\ls{star}^\mu \ktilde\us{emi}_\mu
\\
 \nu\ls{obs}(t\ls{obs}) &\defeq
 - u\ls{obs}^\mu k\us{obs}_\mu =
 - \sigma_c^{-1} u\ls{obs}^\mu \ktilde\us{obs}_\mu
 \,,
\end{split}
\eqe
where $u\ls{star}^\mu$ is the four velocity of S0-2 at $t\ls{emi}$, and $u\ls{obs}^\mu$ is the four velocity  of our observer at $t\ls{obs}$,  $\ktilde\us{emi}_\mu$ and $\ktilde\us{obs}_\mu$ are respectively the 1-from conjugate to the photon's four velocity at the emission event and that at the observation event. 
It is obvious that the definition of redshift \eqref{eq:obs-redshift-def} is not affected by the value of $\sigma_c$.

The expansion of $\rs$ by the parameter $\varepsilon = m/r$ is given by the expansion of frequencies $\nu\ls{emi}$ and $\nu\ls{obs}$. 
Let us calculate the expansion of $\nu\ls{emi}$ from the following form,
\eqb
\label{eq:obs-emissionfreq-primitive}
 \nu\ls{emi} =
 - g\ls{emi}^{\mu\nu} u\us{star}_\mu k\us{emi}_\nu =
 - g\ls{emi}^{00}Ew + g^{0j}\,(w u\us{star}_j + E k\us{emi}_j)
 - g\ls{emi}^{ij} u\us{star}_i k\us{emi}_j
 \,,
\eqe
where $g^{\mu\nu}\ls{emi}$ is the metric at the emission event. 
We need not only the expansion of spatial components $u\us{emi}_j$ and $k\us{emi}_j$ but also the expansion of the conserved quantities $E = -u\us{star}_0$ and $w = - k\us{emi}_0$. 
Further, the expansion of $E$ and $w$ is obtained from the normalization conditions 
$g\ls{emi}^{\mu\nu} u\us{star}_\mu u\us{star}_\nu = -1$ and 
$g\ls{emi}^{\mu\nu} k\us{emi}_\mu k\us{emi}_\nu = 0$. 
Substituting the expansion of $g\ls{emi}^{\mu\nu}$ given in Eq.\eqref{eq:ppn-1PN1PM-inverse} into the normalization conditions, we obtain
\footnote{
The normalization conditions $u^2 = -1$ and $k^2 = 0$ give quadratic equations of $E$ and $w$. 
We choose the solutions satisfying $E>0$ and $w>0$ at the limit of no black hole, $m \to 0$ and $a \to 0$. 
} 
\eqb
\label{eq:obs-Ew-expansion}
\begin{split}
 E &=
 1 + \dfrac{1}{2} \sum_{j=1}^3 (u\us{emi}_j)^2 - \varepsilon\ls{emi}
 + O(\varepsilon\ls{emi}^2)
\\
 w &=
 |k\us{emi}|\,\Bigl[\,
 1 - \Bigl\{\,
      1 + B \Bigl(\dfrac{k\us{emi}\ls{r}}{|k\us{emi}|}\Bigr)^2
      \,\Bigr\}\, \varepsilon\ls{emi}
 + O(\varepsilon\ls{emi}^2)
 \,\Bigr]
 \,,
\end{split}
\eqe
where $\varepsilon\ls{emi} = m/r\ls{emi}$\,, $|k\us{emi}| = \sqrt{\sum_{j=1}^3 (k\us{emi}_j)^2}$ and $\kr\us{emi}$ is in Eq.\eqref{eq:ppn-kr}. 
Then, substituting the expansion of $g\ls{emi}^{\mu\nu}$ and Eq.\eqref{eq:obs-Ew-expansion} into Eq.\eqref{eq:obs-emissionfreq-primitive}, we obtain
\eqb
\label{eq:obs-emissionfreq}
\begin{array}{rll}
 \dfrac{\nu\ls{emi}}{w} =& 1
 &\cdots O(1)
\\
 & - \sum_j \dfrac{k\us{emi}_j}{|k\us{emi}|} u\us{emi}_j
 &\cdots O(\varepsilon\ls{emi}^{0.5})
\\
 & + \dfrac{1}{2}\sum_j (u\us{emi}_j)^2 + \varepsilon\ls{emi}
 &\cdots O(\varepsilon\ls{emi})
\\
 &
 + \Bigl[\,
     \Bigl(\, 1 + B \Bigl(\dfrac{k\us{emi}\ls{r}}{|k\us{emi}|}\Bigr)^2  \,\Bigr)
     \sum_j \dfrac{k\us{emi}_j}{|k\us{emi}|} u\us{emi}_j
   + 2 B \dfrac{k\us{emi}\ls{r}}{|k\us{emi}|} u\us{emi}\ls{r}
   \,\Bigr]\,\varepsilon\ls{emi}
 &\cdots O(\varepsilon\ls{emi}^{1.5})
\\
 & + O(\varepsilon\ls{emi}^2) \,, &
\end{array}
\eqe
where let us note that the expansions of the spatial components $u\us{emi}_j$ and $k\us{emi}_j$ have not been substituted yet, and the order of terms is counted with $k\us{emi}_j \sim O(1)$ and $u\us{emi}_j \sim O(\varepsilon\ls{emi}^{0.5})$.

Next, in order to calculate the expansion of $\nu\ls{obs}$, let us specify the four velocity of our observer, 
\eqb
 u\ls{obs}^\mu = 
 \gamma\ls{obs}
 \left(\,
 1\,,\, v\ls{obs}^x \,,\, v\ls{obs}^y \,,\, v\ls{obs}^z
 \,\right)
 \quad,\quad
 \gamma\ls{obs} =
 \bigl(1-\vec{v}\ls{obs}^{\,2}\bigr)^{-1/2}
 \approx 1
 \,,
\eqe
where $v\ls{obs}^j$ is the spatial velocity in the Cartesian-like coordinates $(x,y,z)$, and we can approximate the gamma factor $\gamma\ls{obs}$ being unity as discussed in Sect.\ref{subsubsec:ppn-setup-XYZ}. 
Then, following the same line of calculations for Eq.\eqref{eq:obs-emissionfreq} together with $u\us{obs}_\mu$, the expansion of $\nu\ls{obs}$ is obtained,
\eqb
\label{eq:obs-observationfreq}
\begin{array}{rcll}
 \dfrac{\nu\ls{obs}}{w} &=& 1
 &\cdots O(1)
\\
 && - \sum_j \dfrac{k\us{obs}_j}{|k\us{obs}|} u\us{obs}_j
 &\cdots O(\varepsilon\ls{obs}^{0.5})
\\
 && + O(\varepsilon\ls{obs}) \,, &
\end{array}
\eqe
where $\varepsilon\ls{obs} = m/r\ls{obs}$, and the expansion of the spatial component $k\us{emi}_j$ has not been substituted yet. 
Further let us note that, due to the order of parameters $\varepsilon\ls{obs} \simeq \vec{v}\ls{obs}^{\,2} \sim 10^{-8}$ as given in Sect.\ref{subsubsec:ppn-setup-XYZ} and $\varepsilon\ls{emi} \simeq \varepsilon\ls{peri} \sim 10^{-3}$ as given in Eq.\eqref{eq:ppn-pnparameter-peri}, we need the expansion of $\nu\ls{obs}$ up to the term of $O(\varepsilon\ls{obs}^{0.5}) \simeq O(\varepsilon\ls{emi}^{1.5})$.

From the above we obtain the expansion of $\rs$ by substituting Eqs.\eqref{eq:obs-emissionfreq} and \eqref{eq:obs-observationfreq} into the definition \eqref{eq:obs-redshift-def}. 
Further we introduce the PN/PM expansion of $u\us{emi}_j$, $k\us{emi}_j$ and $k\us{obs}_j$, which can be expressed as
\eqb
\label{eq:obs-uk-expansion}
 u_j = u\us{(0.5)}_j + u\uone_j + u\us{(1.5)} + O(\varepsilon^2)
 \quad,\quad
 k_j = k\uzero_j + k\uone_j + O(\varepsilon^2)
 \,,
\eqe
where $O(u\us{(n)}_j) \sim \varepsilon^n$ ($n = 0.5, 1, 1.5, \cdots$) due to $\varepsilon \sim \vec{u}^{\,2}$, and $O(k\us{(l)}_j) \sim \varepsilon^l$ ($l = 0,1,2,\cdots$). 
Thus we obtain
\seqb
\label{eq:obs-redshift}
\eqb
 \rs =
 \rs\us{(newton)} + \rs\us{(1PN)} + \rs\us{(1.5PN+1PM)}
 + O(\varepsilon\ls{emi}^2)
 \,,
\eqe
where $\rs\us{(newton)}$ consists of the terms of $O(\varepsilon\ls{emi}^{0.5})$ and $O(\varepsilon\ls{obs}^{0.5})$ which correspond to the formula of redshift in Newtonian dynamics,
\eqb
 \rs\us{(newton)}
 =
   \sum_j \dfrac{k\us{obs(0)}_j}{|k\us{obs(0)}|} v\ls{obs}^j
  - \sum_j \dfrac{k\us{emi(0)}_j}{|k\us{emi(0)}|} u\us{emi(0.5)}_j
 \,,
\eqe
$\rs\us{(1PN)}$ consists of the terms of $O(\varepsilon\ls{emi})$ which include up to 1PN effect of S0-2's motion and 0PM effect of $k\us{emi}_\mu$,
\eqb
 \rs\us{(1PN)}
 =
 \dfrac{1}{2}\sum_j (u\us{emi(0.5)}_j)^2 + \varepsilon\ls{emi}
 - \sum_j \dfrac{k\us{emi(0)}_j}{|k\us{emi(0)}|} u\us{emi(1)}_j
\eqe
and $\rs\us{(1.5PN+1PM)}$ consists of the terms of $O(\varepsilon\ls{emi}^{1.5})$
which include up to the 1.5PN effect of S0-2's motion and the 1PM effect of $k\us{emi}_\mu$,
\eqb
\label{eq:obs-redshift-1.5PN+1PM}
\begin{split}
 \rs\us{(1.5PN+1PM)}
 =&
 \Biggl[\,
 \Bigl(\, 1 + B \Bigl(\dfrac{k\us{emi(0)}\ls{r}}{|k\us{emi(0)}|}\Bigr)^2  \,\Bigr)
 \sum_j \dfrac{k\us{emi(0)}_j}{|k\us{emi(0)}|} u\us{emi(0.5)}_j
 + 2 B \dfrac{k\us{emi(0)}\ls{r}}{|k\us{emi(0)}|} u\us{emi(0.5)}\ls{r}
 \,\Biggr]\,\varepsilon\ls{emi}
\\
 &
 - \sum_j
 \Biggl[\,
 \dfrac{k\us{emi(1)}_j}{|k\us{emi(0)}|} u\us{emi(0.5)}_j
 + \dfrac{k\us{emi(0)}_j \sum_q k\us{emi(1)}_q}{|k\us{emi(0)}|^3} u\us{emi(0.5)}_j
 + \dfrac{k\us{emi(0)}_j}{|k\us{emi(0)}|} u\us{emi(1.5)}_j
 \,\Biggr]
 \,.
\end{split}
\eqe
\seqe
Let us note on the term $\rs\us{(1.5PN+1PM)}$ that the 1.5PN effect of S0-2's motion appears as $u\us{emi (1.5)}_j$ in the last term in Eq.\eqref{eq:obs-redshift-1.5PN+1PM}, and the PPN parameter $B$ does not couple with $u\us{emi (1.5)}_j$. 
The parameter $B$ couples with 0PN effect of S0-2's motion and 0PM effect of photon's emission momentum.

\subsection{Observable PN/PM effects}
\label{subsec:obs-PNPM}

In order to judge the highest PN/PM order which is detectable with the present telescopes, we need typical observational uncertainties of observables,
\seqb
\label{eq:obs-uncertainty}
\eqb
\label{eq:obs-uncertainty-all}
\begin{split}
 \text{obs. uncertainty in astrometry} &:
 \uncertain{\ddec}\,,\, \uncertain{\dra} \sim 10^{-4} \,\,\unit{arcsec}
\\
 \text{obs. uncertainty in spectroscopy} &:
 \uncertain{\rs} \sim 6\times 10^{-5}
 \,\Leftrightarrow\,
 c\,\uncertain{\rs} \sim 20 \,\unit{km/s}
\\
 \text{obs. uncertainty of observation time} &:
 \uncertain{t\ls{obs}} \sim 1 \,\, \unit{day}
 \sim 1.5 \times 10^3 \,\, \unit{min}
 \,,
\end{split}
\eqe
where $\uncertain{O}$ denotes the observational uncertainty of observable $O$. 
Note that, according to the observational values of $\{\ddec,\dra\}$ shown in Appendix~\ref{app:data}, the observational uncertainty in astrometry is typically translated
to
\eqb
\label{eq:obs-uncertainty-astrometry}
 \uncertain{\ddec}\,,\, \uncertain{\dra} \sim 1\%
 \,.
\eqe
\seqe
On the other hand, we find from Eq.\eqref{eq:ppn-value},
\eqb
 O\left(\dfrac{r\ls{emi}}{r\ls{obs}}\right)
 \sim \dfrac{120\,\,\unit{AU}}{8 \,\,\unit{kpc}}
 \sim 10^{-7}
 \,.
\eqe
Further we find the following order relations from Eqs.\eqref{eq:obs-0PM-sol} and \eqref{eq:obs-1PM-sol-b0q0},
\seqb
\label{eq:obs-0PM-order}
\eqb
\begin{split}
 O(\ktilde\us{(0)}_j) &\simeq
 O(\wtilde\us{(0)}) \simeq
 O({\mathcal D}\vec{x}) \simeq
 O(r\ls{obs})
\\
 O(b\uzero) &\simeq O(r\ls{emi} r\ls{obs})
\\
 O(q\uzero_j) &\simeq O(r\ls{emi})
 \,,
\end{split}
\eqe
and from \eqref{eq:obs-1PM-sol},
\eqb
\label{eq:obs-1PM-order}
 O(\ktilde\us{emi (1)}_j) \simeq O(\ktilde\us{obs (1)}_j) \simeq
 O(\wtilde\us{(1)}) \simeq
 O(B \varepsilon\ls{emi} r\ls{obs})
 \,.
\eqe
\seqe
Then we obtain from Eq.\eqref{eq:obs-astrometry},
\eqb
\label{eq:obs-astrometry-uncertainty}
 \left.
 \begin{array}{l}
 O\Bigl(\dfrac{\ddec\us{(1)}}{\ddec\us{(0)}}\Bigr)
 \\
 O\Bigl(\dfrac{\dra\us{(1)}}{\dra\us{(0)}}\Bigr)
 \end{array}
 \right\} 
 \simeq O\left(\dfrac{\ktilde\us{obs (1)}_j}{\ktilde\us{(0)}_j}\right)
 \simeq O(B \varepsilon\ls{emi})
 \simeq B \times 10^{-3}
 \,,
\eqe
and from Eq.\eqref{eq:obs-redshift},
\eqb
\label{eq:obs-redshift-uncertainty}
\begin{split}
 O(c \rs\us{(newton)}) &\simeq O(c\varepsilon\ls{emi}^{0.5}) \simeq 10^3 \,\, \unit{km/s}
\\
 O(c \rs\us{(1PN)}) &\simeq O(c\varepsilon\ls{emi}) \simeq 10^2 \,\, \unit{km/s}
\\
 O(c \rs\us{(1.5PN+1PM)}) &\simeq O(B c \varepsilon\ls{emi}^{1.5})
 \simeq B \,\, \unit{km/s}
 \,,
\end{split}
\eqe
where $\varepsilon\ls{emi} \sim 10^{-3}$ as given in Eq.\eqref{eq:ppn-pnparameter-peri}. 
Further, for the 1PM correction of the observational time \eqref{eq:obs-1PM-sol-tobs1},
we find the following order relation,
\eqb
\label{eq:obs-obstime-uncertainty}
 O(t\ls{obs (1)}) \simeq O(B m) \sim B \dfrac{GM}{c^3} \sim 10 B \,\, \unit{min}
 \,.
\eqe

From Eqs.\eqref{eq:obs-uncertainty} and \eqref{eq:obs-redshift-uncertainty}, the redshift up to the term $\rs\us{(1PN)}$ is already detectable by the present telescope. 
Thus we focus on the detectablity of $\ddec\us{(1)}$, $\dra\us{(1)}$ and $\rs\us{(1.5PN+1PM)}$. 
Comparing the astrometric correction \eqref{eq:obs-astrometry-uncertainty} with $\uncertain{\ddec}$ and $\uncertain{\dra}$ in Eq.\eqref{eq:obs-uncertainty-astrometry}, and the order of $\rs\us{(1.5PN+1PM)}$ in \eqref{eq:obs-redshift-uncertainty} with $\uncertain{\rs}$ in Eq.\eqref{eq:obs-uncertainty-all}, following relations hold,
\seqb
\label{eq:obs-estimation}
\eqb
\label{eq:obs-estimation-astrospectro}
 O(B) \gtrsim 10 \quad\Rightarrow\quad
 \begin{cases}
 O(\ddec\us{(1)}/\ddec\us{(0)}) &\gtrsim 10^{-2} \sim \uncertain{\ddec}
 \\
 O(\dra\us{(1)}/\dra\us{(0)}) &\gtrsim 10^{-2} \sim \uncertain{\dra}
 \\
 O(c \rs\us{(1.5PN+1PM)}) &\gtrsim 10\,\,\unit{km/s} \sim c\uncertain{\rs}
 \,.
 \end{cases}
\eqe
Further, comparing the temporal correction \eqref{eq:obs-obstime-uncertainty} with $\uncertain{t\ls{obs}}$ in Eq.\eqref{eq:obs-uncertainty-all}, we find following relation,
\eqb
\label{eq:obs-estimation-time}
 O(B) \gtrsim 100 \quad\Rightarrow\quad
 O(t\ls{obs(1)}) \gtrsim 10^3 \,\,\unit{min} \sim \uncertain{t\ls{obs}}
 \,.
\eqe
\seqe
From the above estimations, we find some indications for fitting theoretical predictions with observational data.
\begin{itemize}
\item[(i)]
Because the redshift up to $\rs\us{(1PN)}$ is detectable, we must solve the E.O.M of S0-2 \eqref{eq:ppn-timelikegeodesic} at least up to 1PN terms which include the PPN parameters $A$ and $B$ but not the BH's spin effect.
\item[(ii)]
Eq.\eqref{eq:obs-estimation-astrospectro} denotes that, in order to assess whether the case $O(B) \gtrsim 10$ is allowed by the present observational data, we need to calculate the astrometric observables up to $\ddec\us{(1)}$ and $\dra\us{(1)}$, and the redshift up to the terms in $\rs\us{(1.5PN+1PM)}$ depending on $B$. 
This is consistent with the note (i).
\item[(iii)]
From the note (ii), we must calculate the photon's momentum up to 1PM terms $\ktilde\us{emi (1)}_j$ and $\ktilde\us{obs(1)}$, where $\ktilde\us{emi(1)}$ is necessary to the terms in $\rs\us{(1.5PN+1PM)}$ depending on $B$ and $\ktilde\us{obs(1)}$ is necessary to $\ddec\us{(1)}$ and $\dra\us{(1)}$.
\item[(iv)]
Eq.\eqref{eq:obs-estimation-time} denotes that, in order to assess whether the case $O(B) \gtrsim 100$ is allowed by the present observational data, we need to calculate the observational time up to $t\ls{obs(1)}$. 
\item[(v)]
If the true value of $B$ satisfies $O(B) \gtrsim 10$, then it is expected that the fitting of PPN model predictions with observational data can determine the value of $B$ with a sufficiently small fitting error of $B$. 
On the other hand, if the true value of $B$ is of the order of $O(B) \ll 10$, then the fitting result should give a large fitting uncertainty and we can not judge which of PPN model or Schwarzschld case is preferable.
\end{itemize}

\section{$\chi^2$ fitting}
\label{sec:fitting}

As mentioned in the third paragraph of Sect.\ref{sec:intro}, the observations of S0-2's motion have been performed by European group using mainly Very Large Telescope (VLT), American group using mainly Keck telescope, and our Japanese group using mainly Subaru telescope. 
American and European groups have been performing both of the astrometric and spectroscopic observations since 1990s. 
Our Japanese group, since 2014, have been focusing on higher precision spectroscopic observation than the other groups, while a much more time and efforts are required for analyzing raw data. 
The observed values of $\{ \ddec , \dra , \rs \}$ used in this paper are those used in the previous papers by European group \cite{ref:gillessen+2017}, American group \cite{ref:do+2019} and our group \cite{ref:saida+2019}. 
Note that the units of those observational values are usually arcsecond (abbreviated as arcsec) for the astrometric observables and km/s for the spectroscopic observable. 
The summary of those observational values are in Appendix~\ref{app:data}.
(In the European group's paper published in 2020 \cite{ref:gravity2020}, their observational values are not written although some graphs including those data are shown. 
Therefore, we refer their paper published in 2017 \cite{ref:gillessen+2017}, whose observational values are available from their web cite.)

As explained in Sect.\ref{subsubsec:ppn-setup-astrometry}, in comparing theoretical predictions of astrometric observables $\{ \ddec(t),\dra(t) \}$ with observational data, the offsets of the astrometric origins from \sgra given in Eq.\eqref{eq:ppn-astrometry} have to be added to observational data, because the actual astrometric data express the offsets of S0-2's declination and right ascension from the astrometric origins.

Then, we have performed the $\chi^2$ fitting of our PPN predictions of $\{ \ddec, \dra, \rs \}$ with the observational data in Appendix \ref{app:data}. 
Namely, we obtained the values of parameters in Table~\ref{table:parameters} under the condition \eqref{eq:ppn-spin-undetectability} so as to minimize the so-called reduced chi-squared $\chi\ls{red}^2$~\cite{ref:press+1992},
\seqb
\label{eq:fitting-chi2}
\eqb
\begin{split}
 \chi\ls{red}^2
 \,\defeq\,
 \dfrac{1}{N\ls{red}}
 \Biggl[\,
 &
 \sum_{n=1}^{D\ls{Keck}}
 \bigl( \chi\ls{(X.Keck)\it n}^2 + \chi\ls{(Y.Keck)\it n}^2 \bigr)
\\
 &
 +
 \sum_{n=1}^{D\ls{VLT}}
 \bigl( \chi\ls{(X.VLT)\it n}^2 + \chi\ls{(Y.VLT)\it n}^2 \bigr)
 +
 \sum_{n=1}^{D\ls{rs}}
  \chi\ls{(rs)\it n}^2
 \,\Biggr]
 \,,
\end{split}
\eqe
where $D\ls{Keck} (=46)$ is the number of astrometric data taken by American group, $D\ls{VLT} (=144)$ is the number of astrometric data taken by European group, $D\ls{rs} (=123)$ is the number of all spectroscopic data taken by all three groups, and $N\ls{red} = 2 D\ls{Keck} + 2 D\ls{VLT}+D\ls{rs} - 21$ where $21$ is the number of parameters whose values are to be determined by the present fitting process. 
Further the following formulas are used in each term of $\chi\ls{red}^2$, where the terms of astrometric observables of American group are
\eqb
\begin{split}
 \chi\ls{(X.Keck)\it n}^2
 &=
 \Bigl(
 \dfrac{T\ls{ang} \ddec(t_n) - (\dec\us{Keck}_n + O^X\ls{Keck}(t_n)\,)}
 {\uncertain{\dec\us{Keck}_n}}
 \Bigr)^2
\\
 \chi\ls{(Y.Keck)\it n}^2
 &=
 \Bigl(
 \dfrac{T\ls{ang} \dra(t_n) - (\ra\us{Keck}_n + O^Y\ls{Keck}(t_n)\,)}
 {\uncertain{\ra\us{Keck}_n}}
 \Bigr)^2
 \,,
\end{split}
\eqe
the terms of astrometric observables of European group are
\eqb
\begin{split}
 \chi\ls{(X.VLT)\it n}^2
 &=
 \Bigl(
 \dfrac{T\ls{ang} \ddec(t_n) - (\dec\us{VLT}_n + O^X\ls{VLT}(t_n)\,)}
 {\uncertain{\dec\us{VLT}_n}}
 \Bigr)^2
\\
 \chi\ls{(Y.VLT)\it n}^2
 &=
 \Bigl(
 \dfrac{T\ls{ang} \dra(t_n) - (\ra\us{VLT}_n + O^Y\ls{VLT}(t_n)\,)}
 {\uncertain{\ra\us{VLT}_n}}
 \Bigr)^2
 \,,
\end{split}
\eqe
and the term of spectroscopic observable of all groups is
\eqb
 \chi\ls{(rs)\it n}^2
 =
 \Bigl(
 \dfrac{c \rs(t_n) - \rv_n}{\uncertain{\rv_n}}
 \Bigr)^2
 \,,
\eqe
\seqe
where $T\ls{ang} = (180/\pi)\times 60 \times 60$ is the coefficient to change the unit of angle from radian to arcsec, 
the set of values $\{\,(\dec^i_n , \uncertain{\dec^i_n}) \,,\,
(\ra^i_n , \uncertain{\ra^i_n})\,\}$ 
denotes the $n$-th astrometric observational values and its observational uncertainties in the unit of arcsec of American group for $i =$ Keck and those of European group for $i =$ VLT, 
the set of values $(\rv_n , \uncertain{\rv_n})$ denotes the $n$-th spectroscopic observational value and its observational uncertainty in the unit of velocity km/s of all three groups, 
$\vec{O}_i(t_n)$ is the offset of the astrometric origin from \sgra given in Eq.\eqref{eq:ppn-astrometry} at a given observational time $t_n$, and 
$\{ \ddec(t_n) , \dra(t_n) , \rs(t_n) \}$ 
are the PPN model predictions of three observables at $t_n$.

The minimum value of $\chi\ls{red}^2$ is given by the best-fitting parameter values. 
According to the statistics of the co-called $\chi^2$ distribution, the minimum value of $\chi\ls{red}^2$ tends to be unity if the observational data do not contradict the theoretical prediction which is assumed to be consistent with the data.

\begin{table}[!h]
\caption{Best-fitting parameter values, obtained by $\chi^2$ fitting of PPN model prediction with observational data. 
Four parameters $\{a\,,\,\Theta\ls{BH}\,,\,\Phi\ls{BH}\,,\,C_z\}$ are omitted due to the condition \eqref{eq:ppn-spin-undetectability}.
}
\label{table:bestfitting-ppn}
\centering
\begin{tabular}{rcrll}
\hline\hline
\multicolumn{5}{c}{reduced chi-squared for PPN model}
\\
\hline
Minimum of $\chi\ls{red}^2$ &:&
$1.302$ &
&
no dimension
\\[3mm]
(parameter) &&
(best-fitting) &
(fitting error) &
(unit)
\\
\hline\hline
\multicolumn{5}{c}{parameters for BH/\sgra}
\\
\hline
Black hole mass $m$ &:&
$3.9955$ &
$\pm 0.0049$ &
$10^6 M_{\odot}$
\\
PPN parameters $A$ &:&
$22.7$ &
$\pm 1.3$ &
no dimension
\\
$B$ &:&
$-6.92$ &
$\pm 0.93$ &
no dimension
\\[3mm]
\hline\hline
\multicolumn{5}{c}{parameters for observer}
\\
\hline
Distance to \sgra\,  $R\ls{GC}$ &:&
$7.9878$ &
$\pm 0.0043$ &
kpc
\\
Ovserver's velocity $V^X\ls{obs}$ &:&
$0.080$ &
$\pm 0.012$ &
$10^{-3}$ arcsec/yr
\\
$V^Y\ls{obs}$ &:&
$-0.126$ &
$\pm 0.027$ &
$10^{-3}$ arcsec/yr
\\
$V^Z\ls{obs}$ &:&
$-9.61$ &
$\pm 0.49$ &
km/s
\\[3mm]
\hline\hline
\multicolumn{5}{c}{parameters for astrometric origin}
\\
\hline
Keck's astrometry $V^X\ls{Keck}$ &:&
$-0.291$ &
$\pm 0.017$ &
$10^{-3}$ arcsec/yr
\\
$V^Y\ls{Keck}$ &:&
$0.016$ &
$\pm 0.029$ &
$10^{-3}$ arcsec/yr
\\
$A^X\ls{Keck}$ &:&
$-1.522$ &
$\pm 0.091$ &
$10^{-3}$ arcsec
\\
$A^Y\ls{Keck}$ &:&
$0.928$ &
$\pm 0.082$ &
$10^{-3}$ arcsec
\\
VLT's astrometry $V^X\ls{VLT}$ &:&
$-0.259$ &
$\pm 0.017$ &
$10^{-3}$ arcsec/yr
\\
$V^Y\ls{VLT}$ &:&
$-0.003$ &
$\pm 0.029$ &
$10^{-3}$ arcsec/yr
\\
$A^X\ls{VLT}$ &:&
$0.033$ &
$\pm 0.091$ &
$10^{-3}$ arcsec
\\
$A^Y\ls{VLT}$ &:&
$-0.728$ &
$\pm 0.084$ &
$10^{-3}$ arcsec
\\[3mm]
\hline\hline
\multicolumn{5}{c}{parameters for S0-2's initial condition}
\\
\hline
Apocenter observation $t\ls{obs.apo}$ &:&
$2010.335099$ &
$\pm 0.000026$ &
AD
\\
Orbital period $T\ls{star}$ &:&
$16.06061$ &
$\pm 0.00028$ &
yr
\\
Orbital eccentricity $e\ls{star}$ &:&
$0.885051$ &
$\pm 0.000030$ &
no dimension
\\
Inclination angle $I\ls{star}$ &:&
$133.960$ &
$\pm 0.016$ &
degree
\\
Ascending node angle $\Omega\ls{star}$ &:&
$227.809$ &
$\pm 0.026$ &
degree
\\
Pericenter angle $\omega\ls{star}$ &:&
$66.339$ &
$\pm 0.018$ &
degree
\\
\hline
\end{tabular}
\end{table}

Our fitting result of the PPN model with the observational data is summarized in Table~\ref{table:bestfitting-ppn}. 
We performed simulations for the $\chi^2$ fitting with Mathematica. 
The fitting method is a simple minimum search of $\chi\ls{red}^2$, and we have stopped the minimum search when the improvement of $\chi\ls{red}^2$ becomes less than $10^{-6}$. 
The fitting error in Table~\ref{table:bestfitting-ppn} is calculated from the covariance matrix $C_{IJ}$~\cite{ref:press+1992},
\seqb
\label{eq:fitting-error}
\eqb
 C_{IJ} \defeq
 \dfrac{1}{2}
 \Biggl[
 \pd{^2 (N\ls{red} \chi\ls{red}^2)}
 {I \partial J}
 \Biggr]^{-1}
 \,,
\eqe
where $I$ and $J$ are the indices denoting the 21 parameters $I, J = m \,,\, A \,,\, B \,,\, R\ls{GC} \,,\, \cdots$, the power $-1$ in the right hand side denotes the inverse matrix, and the fitting error $\uncertain{J}$ of a parameter $J$ is given by
\eqb
 \uncertain{J} \defeq \sqrt{C_{JJ}}\, \bigr|_\text{best-fitting}
 \,.
\eqe
\seqe

Further, in Table~\ref{table:bestfitting-sch}, the result of $\chi^2$ fitting of the Schwarzschild case with the observational data is summarized, where the Schwarzschild case is given by fixing PPN parameters at $\{A,B\} = \{0,1\}$.

\begin{table}[!h]
\caption{Best-fitting parameter values, obtained by $\chi^2$ fitting of Schwarzschild model prediction with observational data. 
The PPN parameters are fixed at $\{A\,,\,B\} = \{ 0 \,,\, 1 \}$, and four parameters $\{a\,,\,\Theta\ls{BH}\,,\,\Phi\ls{BH}\,,\,C_z\}$ are omitted due to the condition \eqref{eq:ppn-spin-undetectability}.
}
\label{table:bestfitting-sch}
\centering
\begin{tabular}{rcrll}
\hline\hline
\multicolumn{5}{c}{reduced chi-squared for Schwarzschild model}
\\
\hline
Minimum of $\chi\ls{red}^2$ &:&
$1.318$ &
&
no dimension
\\[3mm]
(parameter) &&
(best-fitting) &
(fitting error) &
(unit)
\\
\hline\hline
\multicolumn{5}{c}{parameter for BH/\sgra}
\\
\hline
Black hole mass $m$ &:&
$4.017$ &
$\pm 0.038$ &
$10^6 M_{\odot}$
\\[3mm]
\hline\hline
\multicolumn{5}{c}{parameters for observer}
\\
\hline
Distance to \sgra\,  $R\ls{GC}$ &:&
$8.008$ &
$\pm 0.037$ &
kpc
\\
Ovserver's velocity $V^X\ls{obs}$ &:&
$0.084$ &
$\pm 15.088$ &
$10^{-3}$ arcsec/yr
\\
$V^Y\ls{obs}$ &:&
$-0.130$ &
$\pm 13.483$ &
$10^{-3}$ arcsec/yr
\\
$V^Z\ls{obs}$ &:&
$-11.26$ &
$\pm 2.46$ &
km/s
\\[3mm]
\hline\hline
\multicolumn{5}{c}{parameters for astrometric origin}
\\
\hline
Keck's astrometry $V^X\ls{Keck}$ &:&
$-0.289$ &
$\pm 15.084$ &
$10^{-3}$ arcsec/yr
\\
$V^Y\ls{Keck}$ &:&
$0.038$ &
$\pm 13.480$ &
$10^{-3}$ arcsec/yr
\\
$A^X\ls{Keck}$ &:&
$-1.623$ &
$\pm 0.148$ &
$10^{-3}$ arcsec
\\
$A^Y\ls{Keck}$ &:&
$0.793$ &
$\pm 0.127$ &
$10^{-3}$ arcsec
\\
VLT's astrometry $V^X\ls{VLT}$ &:&
$-0.263$ &
$\pm 15.090$ &
$10^{-3}$ arcsec/yr
\\
$V^Y\ls{VLT}$ &:&
$0.015$ &
$\pm 13.487$&
$10^{-3}$ arcsec/yr
\\
$A^X\ls{VLT}$ &:&
$-0.061$ &
$\pm 0.150$ &
$10^{-3}$ arcsec
\\
$A^Y\ls{VLT}$ &:&
$-0.837$ &
$\pm 0.127$ &
$10^{-3}$ arcsec
\\[3mm]
\hline\hline
\multicolumn{5}{c}{parameters for S0-2's initial condition}
\\
\hline
Apocenter observation $t\ls{obs.apo}$ &:&
$2010.33573$ &
$\pm 0.00072$ &
AD
\\
Orbital period $T\ls{star}$ &:&
$16.0487$ &
$\pm 0.0013$ &
yr
\\
Orbital eccentricity $e\ls{star}$ &:&
$0.88558$ &
$\pm 0.00032$ &
no dimension
\\
Inclination angle $I\ls{star}$ &:&
$134.01$ &
$\pm 0.12$ &
degree
\\
Ascending node angle $\Omega\ls{star}$ &:&
$227.85$ &
$\pm 0.12$ &
degree
\\
Pericenter angle $\omega\ls{star}$ &:&
$66.394$ &
$\pm 0.092$ &
degree
\\
\hline
\end{tabular}
\end{table}

Note that, as explained in Sect.\ref{subsubsec:ppn-setup-IC}, the orbital period $T\ls{star}$ shown in Tables~\ref{table:bestfitting-ppn} and~\ref{table:bestfitting-sch} is the Keplerian approximation given by the initial conditions~\eqref{eq:ppn-IC-rv}. 
On the other hand, in our best-fitting PPN model, the time interval from the apocenter passage in 2010 to the next apocenter passage in 2026 becomes $16.0508$ yr, and the time interval from the pericenter passage in 2018 to the next pericenter passage in 2036 becomes $16.0509$ yr.
These time intervals are different from $T\ls{star}$ in Table~\ref{table:bestfitting-ppn} by $0.01$ yr, a few days. 
This estimation supports the discussion after Eq.\eqref{eq:ppn-IC-rv} that the Keplerian period $T\ls{star}$ is a good approximation as the observational orbital period.

\section{Discussions}
\label{sec:discussions}

Using the observational data of S0-2's motion, we have been performing a PPN test of the black hole metric of \sgra. 
Through formulating the PPN model, we have found a possibility that the gravitational lens effect is detectable under the present observational uncertainties, as estimated in Sect.\ref{subsec:obs-PNPM}. 
This possibility is a new finding by this paper, because this possibility had not been recognized in all previous works by all groups practicing the observation of S0-2's motion~\cite{ref:ghez+2000,ref:schoedel+2002,ref:gillessen+2017,
ref:gravity2018,ref:do+2019,ref:saida+2019,ref:gravity2020}.

The best-fitting values of PPN parameters in Table~\ref{table:bestfitting-ppn} are
\eqb
 \text{Best-fitting PPN parameter}:
 \{A,B\} = \{22.7\pm 1.3 \,,\, -6.92\pm 0.93 \}
 \,.
\eqe 
This result does not include the 
Schwarzschild case $\{A,B\} = \{0,1\}$ within the fitting error. 
This implies that the spacetime of \sgra is not of Schwarzschild metric. 
If this is true, then we need to examine the possibilities (I) and (II) mentioned in Sect.\ref{sec:intro}.

Note that, because the minimum $\chi\ls{red}^2$ for PPN model in Table~\ref{table:bestfitting-ppn} is closer to unity than that for Schwarzschild case in Table~\ref{table:bestfitting-sch}, one may think the PPN model is better than the Schwarzschild case. 
However, the difference between the minimum $\chi\ls{red}^2$ for PPN model and that for Schwarzschild case is of $O(0.01)$, not large enough to ensure the statistical significance of non-Schwarzschild result in the framework of $\chi^2$ statistics. 
If this difference was about or greater than $O(1)$, then the non-Schwarzschild result was statistically significant in the framework of $\chi^2$ statistics.
Therefore, a more precise statistical analysis than $\chi^2$ statistics is necessary to extract a statistically significant information of the gravitational field of \sgra from present observational data. 
We will report a result by a Bayesian analysis in next paper.

Although the statistical significance is not obtained from the values of $\chi\ls{red}^2$, we may have some insights into a statistical discrimination between PPN model and Schwarzschild case from Tables~\ref{table:bestfitting-ppn} and~\ref{table:bestfitting-sch}. 
Let us focus on six parameters 
$\{ V\ls{obs}^X , V\ls{obs}^Y , V\ls{Keck}^X , V\ls{Keck}^Y , V\ls{VLT}^X , V\ls{VLT}^Y \}$, which are the velocities on the 2D sky plane and can not be measured without astrometric observational data. 
The best-fitting values of these six parameters in PPN model and those in Schwarzschild case are similar. 
However, the fitting errors of them for Schwarzschild case are much larger than those for PPN model as shown in Tables~\ref{table:bestfitting-ppn} and~\ref{table:bestfitting-sch}. 
In other words, the best-fitting values of the six parameters in PPN model are determined with much better statistical significance than those in Schwarzschild case.
Thus, it may be expected that the PPN model fits with the present observational data better than Schwarzschild case. 
In order to check whether this insight is true, we are planning to perform a hierarchical Bayesian fitting of PPN model with observational data.

\section*{Acknowledgment}

This research is based in part on data collected at Subaru Telescope operated by the National Astronomical Observatory of Japan.
We are honored and grateful for the opportunity of observing the Galactic center from Maunakea, which has the cultural, historical and natural significance in Hawaii. 
In addition, we would like to express our gratitude to staffs of Subaru telescope, for their continuous supports for our observations over 10 years. 
H.S. was supported by JSPS KAKENHI, Grant-in-Aid for Scientific Research (B) 19H01900. 
S.N. was supported by JSPS KAKENHI, Grant-in-Aid for Scientific Research (A) 19H00695.



\let\doi\relax 


\appendix

\section{PPN/PPM expansion with all PPN parameters $\{A,B,C_\perp,C_z,N\ls{t},N\ls{s}\}$}
\label{app:ppn}

If the black hole mass $m$ and spin $a$ are already known, these parameters $\{m,a\}$ and the PPN parameters ${\bf X}\ls{ppn} = \{A,B,C_\perp,C_z,N\ls{t},N\ls{s}\}$ do not degenerate in the PPN metric \eqref{eq:ppn-all}. 
For this case, the PPN expansion of timelike geodesic equations and the PPM expansion of null geodesic equations become as follows.

The PPN expansion of the Hamiltonian of timelike geodesics \eqref{eq:ppn-hamiltonian-timelike} is
\eqb
\label{eq:app-Hu-expansion}
\begin{array}{rcll}
 \Hu(x,u)
 &=&
 \displaystyle
 - \dfrac{1}{2} E^2 + \dfrac{1}{2} \sum_{j=1}^3 u_j^2
 - N\ls{t} E^2 \varepsilon(r)
 &\cdots\text{0PN : $O(\varepsilon)$}
\\
 &&
 - N\ls{s}\, E\, \ur\, \varepsilon(r)
 &\cdots\text{0.5PN : $O(\varepsilon^{1.5})$}
\\
 &&
 - B \ur^2 \varepsilon(r)
 - \dfrac{4 N\ls{t}^2 + A - N\ls{s}^2}{2} E^2 \varepsilon(r)^2
 &\cdots\text{1PN : $O(\varepsilon^2)$}
\\
 &&
 \!\!
 \begin{array}{l}
 - 2\,\Bigl[
   (N\ls{t}-B) N\ls{s} \ur
 \\
 \phantom{- 2\,\Bigl[}
   -\dfrac{a}{m} \Bigl(\,
   C_\perp \dfrac{L_z}{r} - C_z \dfrac{z}{r} \ur
   \,\Bigr)
 \,\Bigr]\,E\,\varepsilon(r)^2
 \end{array}
 &\cdots\text{1.5PN : $O(\varepsilon^{2.5})$}
\\
 &&
 + O(\varepsilon^3)
 &\cdots\text{higer PN}
 \,.
\end{array}
\eqe
From this Hamiltonian, the PPN timelike geodesic equations corresponding to Eqs.\eqref{eq:ppn-timelikegeodesic} are
\seqb
\label{eq:app-timelikegeodesic}
\eqb
\label{eq:app-timelikegeodesic-xj}
\begin{array}{rcll}
 E \od{x^j(t)}{t}
 &=& E \od{x^j(\tau)}{\tau}\,\Bigl(\od{t(\tau)}{\tau}\Bigr)^{-1}
 &
\\
 &=&
 u_j
 &
 \cdots\text{0PN}
\\
 &&
 - N\ls{s}E\,\dfrac{x^j}{r}\,\varepsilon(r)
 &
 \cdots\text{0.5PN}
\\
 &&
 - 2 B\, \ur \dfrac{x^j}{r} \,\varepsilon(r)
 - 2 N\ls{t}\, u_j \,\varepsilon(r)
 &
 \cdots\text{1PN}
\\
 &&
 \!\!\!
 \left.
 \begin{array}{l}
 - 2\,\Bigl[
   (3N\ls{t}-B) N\ls{s} \dfrac{x^j}{r}
 \\
 \phantom{- 2\,\Bigl[}
   -\dfrac{a}{m} \Bigl(\,
   C_\perp \dfrac{d^j}{r} - C_z \dfrac{z}{r} \dfrac{x^j}{r}
   \,\Bigr)\,\Bigr]\,E\,\varepsilon(r)^2
\\
  -\dfrac{N\ls{s}}{E}\,\ur\,u_j\,\varepsilon(r)
 \end{array}
 \right\}
 &
 \cdots\text{1.5PN}
\\
 && 
 + O(\varepsilon^{2.5})
 &
 \cdots\text{higher PN}
\end{array}
\eqe
\eqb
\label{eq:app-timelikegeodesic-uj}
\begin{array}{rcll}
 E \od{u_j(t)}{t}
 &=&
 E \od{u_j(\tau)}{\tau}\,\Bigl(\od{t(\tau)}{\tau}\Bigr)^{-1}
 &
\\
 &=&
 - N\ls{t} E^2 \, \dfrac{x^j}{r^2}\,\varepsilon(r)
 &
 \cdots\text{0PN}
\\
 &&
 + N\ls{s}E\,\Bigl(\, -2\dfrac{x^j}{r^2} \ur + \dfrac{u_j}{r}
    \,\Bigr)\,\varepsilon(r)
 &
 \cdots\text{0.5PN}
\\
 &&
 \!\!\!
 \left.
 \begin{array}{l}
 + B\,\Bigl( 3 \dfrac{x^j}{r} \ur^2 - 2 u_j \ur \,\Bigr)\,\dfrac{\varepsilon(r)}{r}
 \\
 \qquad + \bigl( 6 N\ls{t}^2 - A + N\ls{s}^2 \bigr) E^2 \dfrac{x^j}{r^2} \varepsilon(r)^2
 \end{array}
 \qquad\right\}
 &
 \cdots\text{1PN}
\\
 &&
 \!\!\!
 \left.
 \begin{array}{l}
  + 6 \widetilde{P}(x,u)\,E\,\dfrac{x^j}{r^2} \varepsilon(r)^2
  - 2 \widetilde{Q}_j(x,u)\,E\,\dfrac{\varepsilon(r)^2}{r}
 \\
  + N\ls{t}N\ls{s}E\,\Bigl(\, 5 \dfrac{x^j}{r^2} \ur - 2 \dfrac{u_j}{r}
     \,\Bigr)\,\varepsilon(r)^2
 \end{array}
 \right\}
 &
 \cdots\text{1.5PN}
\\
 && 
 + O(\varepsilon^3)
 &
 \cdots\text{higer PN}
 \,,
\end{array}
\eqe
where $\ur$ is given in Eq.\eqref{eq:ppn-ur}, $d^j = (-y,x,0)$ in 1.5PN terms of Eq.\eqref{eq:app-timelikegeodesic-xj}, and $\widetilde{P}(x,u)$ and $\widetilde{Q}_j(x,u)$ in 1.5PN terms of Eq.\eqref{eq:app-timelikegeodesic-uj} are
\eqb
\label{eq:app-PQ}
\begin{split}
 \widetilde{P}(x,u) &\defeq
 ( N\ls{t} - B ) N\ls{s} \ur
 - \dfrac{a}{m} \Bigl(\,
 C_\perp \dfrac{L_z}{r} - C_z \dfrac{z}{r} \ur \,\Bigr)
\\
 \widetilde{Q}_j(x,u) &\defeq
 ( N\ls{r} - B ) N\ls{s} \ur
 - \dfrac{a}{m} \Bigl[\, C_\perp q_j
 -C_z \Bigl\{
    \delta_{j3} \ur
    + \dfrac{z}{r}\Bigl(\, \dfrac{u_j}{r} - \dfrac{x^j}{r^2}\ur \,\Bigr)
 \,\Bigr\}\,\Bigr]
 \,,
\end{split}
\eqe
\seqe
where $q_j = (u_2 , -u_1 , 0)$. 
From Eqs.\eqref{eq:app-timelikegeodesic}, it is obvious that the PPN parameters $N\ls{t}$ and $N\ls{s}$ appear, respectively, in the 0PN terms and the 0.5PN terms.

The PPM expansion of the Hamiltonian of null geodesics \eqref{eq:ppn-hamiltonian-null} is
\eqb
\label{eq:app-Hk-expansion}
\begin{array}{rcll}
 \Hk(x,k)
 &=&
 \displaystyle
 \dfrac{1}{2}g^{00} w^2
 - g^{0j}\, k_j \, w
 + \dfrac{1}{2} g^{ij} k_i k_j
 &
\\
 &=&
 \displaystyle
 - \dfrac{1}{2} w^2 + \dfrac{1}{2} \sum_{j=1}^3 k_j^2
 &\cdots\text{0PM : $O(1)$}
\\
 &&
 - \Bigl(
   N\ls{(t)} w^2 + N\ls{(s)} k\ls{r}\,w + B\, k\ls{r}^2
   \,\Bigr)\,\varepsilon(r)
 &\cdots\text{1PM : $O(\varepsilon)$}
\\
 &&
 + O(\varepsilon^2)
 &\cdots\text{higher PM}
 \,.
\end{array}
\eqe
From this Hamiltonian, the PPM null geodesic equations corresponding to Eqs.\eqref{eq:ppn-nullgeodesic} are
\seqb
\label{eq:app-nullgeodesic}
\eqb
\label{eq:app-nullgeodesic-t}
\begin{array}{rcll}
 \od{t(\sigma)}{\sigma} &=&
 \pd{\Hk\bigl(\,x^\alpha(\sigma)\,,\,k_\alpha(\sigma)\,\bigr)}{(-w)} &
\\
 &=&
 w
 &
 \cdots\text{0PM}
\\
 &&
 + \bigl( 2 N\ls{(t)} w + N\ls{(s)} k\ls{r} \bigr)\,\varepsilon(r)
 &
 \cdots\text{1PM}
\\
 && 
 + O(\varepsilon^2)
 &
 \cdots\text{higher PM}
\end{array}
\eqe
\eqb
\label{eq:app-nullgeodesic-xj}
\begin{array}{rcll}
 \od{x^j(\sigma)}{\sigma} &=& \pd{\Hk\bigl(\,x^\alpha(\sigma)\,,\,k_\alpha(\sigma)\,\bigr)}{k_j(\sigma)} &
\\
 &=&
 k_j
 &
 \cdots\text{0PM}
\\
 &&
 - \Bigl[\,
   N\ls{s} w + 2 B k\ls{r}
   \,\Bigr]\,\dfrac{x^j}{r}\varepsilon(r)
 &
 \cdots\text{1PM}
\\
 && 
 + O(\varepsilon^2)
 &
 \cdots\text{higher PM}
\end{array}
\eqe
\eqb
\label{eq:app-nullgeodesic-uj}
\begin{array}{rcll}
 \od{k_j(\sigma)}{\sigma} &=&
 -\pd{\Hk\bigl(\,x^\alpha(\sigma)\,,\,k_\alpha(\sigma)\,\bigr)}{x^j(\sigma)} &
\\
 &=& 0
 &
 \cdots\text{0PM}
\\
 &&
 \!\!\!
 \left.
 \begin{array}{l}
 + \Bigr[\,
 - N\ls{(t)} w^2 \dfrac{x^j}{r}
 \\
 \phantom{ + \Bigr[\,}
 + N\ls{(s)} w\, \Bigl( k_j - 2 \dfrac{x^j}{r} k\ls{r} \Bigr)
 \\
 \phantom{ + \Bigr[\,}
 - B \Bigl( 3\dfrac{x^j}{r} k\ls{r}^2 - 2k_j k\ls{r} \Bigr)
 \,\Bigr]\,\dfrac{\varepsilon(r)}{r}
 \end{array}
 \right\}
 &
 \cdots\text{1PM}
\\
 && 
 + O(\varepsilon^2)
 &
 \cdots\text{higher PM}
 \,,
\end{array}
\eqe
\seqe
where $\kr$ is given in Eq.\eqref{eq:ppn-kr}.

\section{Integral formulas for obtaining 1PM solution \eqref{eq:obs-1PM-sol}}
\label{app:1PM}

In obtaining the 1PM solution \eqref{eq:obs-1PM-sol}, the integral of 0PM radial quantity 
$r\lzero(\sigmatilde) = (x\lzero^2+y\lzero^2+z\lzero^2)^{1/2}$ 
is necessary. 
By substituting the 0PM solution \eqref{eq:obs-0PM-sol} into $r\lzero(\sigmatilde)$, we have
\eqb
 r\lzero(\sigmatilde)^2 =
 \wtilde^{\rm (0)\,2} \,\sigmatilde^2 + 2 b\uzero \sigmatilde + r\ls{emi}^2
 \,,
\eqe
where $\wtilde\uzero$ and $b\uzero$ are respectively in Eq.\eqref{eq:obs-0PM-equation-wtilde0} and Eq.\eqref{eq:obs-1PM-sol-b0q0}, which are independent of $\sigmatilde$. 
Then, when integrating the 1PM differential equations \eqref{eq:obs-1PM-equation-differential}, the basic integral for obtaining the 1PM solution is of the following form,
\eqb
\label{eq:app-integral-basic}
 I[l,n] \defeq
 \int \diff{\sigmatilde}\,\,
 \sigmatilde^{\,l}
 \left(
 \sqrt{\alpha \sigmatilde^2 + \beta \sigmatilde + \gamma}
 \,\right)^n
 \,,
\eqe
where $\alpha = \wtilde^{\rm (0)\,2}$, $\beta = 2 b\uzero$ and $\gamma = r\ls{emi}^2$. 
The following cases of $I[l,n]$ are useful for obtaining the 1PM solution \eqref{eq:obs-1PM-sol}.
\eqb
\label{eq:app-integral-formulas}
\begin{split}
 I[0,-1] &=
 \dfrac{1}{\sqrt{\alpha}}
 \ln\left|\,
 2\alpha \sigmatilde + \beta
 + 2 \sqrt{\alpha (\alpha \sigmatilde^2 + \beta \sigmatilde + \gamma)}
 \,\right|
\\
 I[1,-1] &=
 \dfrac{\sqrt{\alpha \sigmatilde^2 + \beta \sigmatilde + \gamma}}{\alpha}
 - \dfrac{\beta}{2 \alpha} I[0,-1]
\\
 I[0,n] &=
 \dfrac{2 ( 2 \alpha \sigmatilde + \beta)}{(n+2) (\beta^2 - 4 \alpha \gamma)}
 \left( \alpha \sigmatilde^2 + \beta \sigmatilde + \gamma \right)^{1+n/2}
 - \dfrac{4 (n+3) \alpha}{(n+2) (\beta^2 - 4 \alpha \gamma)} I[0,n+2]
\\
 I[1,n] &=
 \dfrac{1}{(n+2) \alpha}
 \left( \alpha \sigmatilde^2 + \beta \sigmatilde + \gamma \right)^{1+n/2}
 - \dfrac{\beta}{2 \alpha} I[0,n]
 \,,
\end{split}
\eqe
where $\alpha > 0$ for $I[0,-1]$ and $n \neq -2$ for $I[1,n]$ and $I[0,n]$.

\section{Observational data}
\label{app:data}

This appendix shows the list of observational data used in Sect.\ref{sec:fitting}. 
Total number of observational data is $503$ being composed of $123$ redshift data and $190$ astrometric data, where one astrometric data includes two data values of right  ascension and declination of S0-2's position on the sky plane. 
We divide these data by grouping the date of observation in order to adjust the table size to the page size. 
Following are the notes for those tables.
\begin{itemize}
\item
The date of observation is the median of the duration of observational operation, and shown with the sideral year, $1\,\,\unit{yr} = 365.25636\,\,\unit{day}$. 
The duration is usually about one day, and the uncertainty of observation time in Eq.\eqref{eq:obs-uncertainty-all} is $\uncertain{t\ls{obs}} \sim 1\,\,\unit{day}$.
Then, due to $0.00636\,\,\unit{day} \ll \uncertain{t\ls{obs}}$, we set $1\,\,\unit{yr} = 365.25\,\,\unit{day}$ in our simulation of $\chi^2$ fitting. 
\item
The unit of spectroscopic observed value $z\ls{rs}$ and uncertainty $\uncertain{z\ls{rs}}$ is km/s.
\item
The unit of astrometric observed value $\{ \ddec,\dra \}$ and uncertainty $\{ \uncertain{\ddec},\uncertain{\dra} \}$ is milli-arcsecond, abbreviated as ``mas''.
\item
In the column of ``obs'' in all tables, the symbols ``A'', ``E'' and ``J'' denote respectively American (Keck), European (VLT) and Japanese (Subaru) groups' observation. 
\end{itemize}
Further, let us make some additional notes to the data set shown below.
\begin{itemize}
\item
American and Japanese groups have released all observed values and uncertainties.
\item
European group had released the observational values and uncertainties until 2016, but have not released those values from 2017.
\item
Any astronomical observation is affected rather strongly by weather conditions, and one may think the variability of observational uncertainties is larger than that of usual ground experiments in Physics. 
\item
The pericenter passage of S0-2 occurred in May 2018. 
However in 2018, a rather big eruption of Kilauea volcano occurred in Hawaii island where the telescopes used by Japanese and American groups are located at the summit of Mt. Maunakea. 
Further a bad weather condition due to ``La Nina'' had continued through 2018. 
Because of these unexpected bad conditions, the number of Japanese data in 2018 were less than our original plan, and the observational uncertainties in 2018 became larger than those in previous data.
\end{itemize}

\begin{table}[!h]
\caption{
Spectroscopic data from 2000 to 2013, the unit is km/s.
}
\label{table:data-spectroscopy-1}
\centering
\begin{tabular}{lrlc|}
date & $z\ls{rs}$ & $\uncertain{z\ls{rs}}$ & obs
\\
\hline
2000.4764 & 1199 & 100 & A
\\
2002.4175 & $-$473 & 39 & A
\\
2002.4203 & $-$476 & 39 & A
\\
2003.2710 & $-$1571 & 59 & E
\\
2003.3530 & $-$1512 & 40 & E
\\
2003.4333 & $-$1593 & 34 & A
\\
2003.4360 & $-$1522 & 36 & A
\\
2003.4460 & $-$1428 & 51 & E
\\
2004.4750 & $-$1149 & 47 & A
\\
2004.5350 & $-$1055 & 46 & E
\\
2004.5370 & $-$1056 & 37 & E
\\
2004.6320 & $-$1039 & 39 & E
\\
2005.1580 & $-$1001 & 77 & E
\\
2005.2120 & $-$960 & 37 & E
\\
2005.2150 & $-$910 & 54 & E
\\
2005.4100 & $-$964 & 37 & A
\\
2005.4550 & $-$839 & 60 & E
\\
2005.4610 & $-$907 & 43 & E
\\
2005.5031 & $-$844 & 18 & A
\\
2005.6770 & $-$774 & 77 & E
\\
2005.7690 & $-$860 & 58 & E
\\
2006.2040 & $-$702 & 42 & E
\\
2006.3050 & $-$718 & 77 & E
\\
2006.4613 & $-$711 & 25 & A
\\
2006.4942 & $-$667 & 25 & A
\\
2006.4969 & $-$688 & 26 & A
\\
2006.6240 & $-$658 & 57 & E
\\
2007.2300 & $-$586 & 57 & E
\\
2007.3040 & $-$537 & 57 & E
\\
2007.5449 & $-$489 & 22 & A
\\
2007.5500 & $-$505 & 57 & E
\\
2007.6730 & $-$482 & 57 & E
\\
2008.2620 & $-$394 & 27 & E
\\
2008.3723 & $-$384 & 18 & A
\\
2008.4310 & $-$425 & 62 & E
\\
2008.5634 & $-$379 & 17 & A
\end{tabular}
\begin{tabular}{|lrlc}
date & $z\ls{rs}$ & $\uncertain{z\ls{rs}}$ & obs
\\
\hline
2009.3415 & $-$254 & 16 & A
\\
2009.3443 & $-$291 & 14 & A
\\
2009.3850 & $-$241 & 45 & E
\\
2010.3491 & $-$146 & 18 & A
\\
2010.3540 & $-$134 & 27 & E
\\
2011.3170 & $-$3 & 34 & E
\\
2011.5204 & 9 & 28 & A
\\
2011.5670 & 35 & 57 & E
\\
2012.2100 & 185 & 34 & E
\\
2012.3420 & 167 & 34 & E
\\
2012.4353 & 156 & 26 & A
\\
2012.4380 & 165 & 23 & A
\\
2012.4435 & 182 & 16 & A
\\
2012.4940 & 195 & 34 & E
\\
2012.5130 & 186 & 34 & E
\\
2012.5525 & 182 & 20 & A
\\
2012.5553 & 191 & 17 & A
\\
2012.6128 & 204 & 15 & A
\\
2012.6154 & 186 & 20 & A
\\
2012.7050 & 190 & 45 & E
\\
2013.2620 & 313 & 23 & E
\\
2013.3580 & 328 & 20 & A
\\
2013.3607 & 330 & 17 & A
\\
2013.3635 & 305 & 20 & A
\\
2013.3662 & 283 & 16 & A
\\
2013.3717 & 326 & 20 & A
\\
2013.3744 & 306 & 21 & A
\\
2013.5628 & 382 & 28 & A
\\
2013.5656 & 347 & 39 & A
\\
2013.5683 & 356 & 16 & A
\\
2013.6065 & 370 & 20 & A
\\
2013.6093 & 352 & 32 & A
\\
2013.6148 & 349 & 16 & A
\\
2013.6550 & 361 & 45 & E
\\
2013.7260 & 384 & 34 & E
\\
\phantom{emptyline}
\end{tabular}
\end{table}

\begin{table}[!h]
\caption{
Spectroscopic data from 2014 to 2018, the unit is km/s.
}
\label{table:data-spectroscopy-2}
\centering
\begin{tabular}{lrlc|}
date & $z\ls{rs}$ & $\uncertain{z\ls{rs}}$ & obs
\\
\hline
2014.1850 & 490 & 28 & E
\\
2014.2630 & 515 & 34 & E
\\
2014.3765 & 488 & 19 & A
\\
2014.379 & 485.6 & 27.3 & J
\\
2014.3901 & 513 & 18 & A
\\
2014.5019 & 545 & 17 & A
\\
2014.5210 & 568 & 17 & E
\\
2015.2990 & 765 & 23 & E
\\
2015.3374 & 740 & 19 & A
\\
2015.5506 & 831 & 17 & A
\\
2015.636 & 886.5 & 17.5 & J
\\
2015.7060 & 869 & 45 & E
\\
2016.2840 & 1081 & 45 & E
\\
2016.3669 & 1104 & 17 & A
\\
2016.3696 & 1145 & 19 & A
\\
2016.3723 & 1140 & 16 & A
\\
2016.381 & 1096.2 & 17.9 & J
\\
2016.5190 & 1198 & 34 & E
\\
2017.343 & 1768.7 & 21.3 & J
\\
2017.348 & 1798.8 & 15.6 & J
\\
2017.3745 & 1809 & 13 & A
\\
2017.3772 & 1807 & 19 & A
\\
2017.3799 & 1782 & 14 & A
\\
2017.5464 & 1998 & 17 & A
\\
2017.5683 & 2014 & 15 & A
\\
2017.605 & 2133.3 & 27.8 & J
\end{tabular}
\begin{tabular}{|lrlc}
date & $z\ls{rs}$ & $\uncertain{z\ls{rs}}$ & obs
\\
\hline
2017.609 & 2169.6 & 37.1 & J
\\
2017.6175 & 2095 & 16 & A
\\
2017.6694 & 2214 & 20 & A
\\
2018.2071 & 3798 & 22 & A
\\
2018.240 & 4001.9 & 37.2 & J
\\
2018.243 & 4096.6 & 40.1 & J
\\
2018.3109 & 3966 & 15 & A
\\
2018.3628 & 3000 & 13 & A
\\
2018.3684 & 2804 & 18 & A
\\
2018.3739 & 2622 & 63 & A
\\
2018.382 & 2466.4 & 67.7 & J
\\
2018.3874 & 2130 & 20 & A
\\
2018.3901 & 2038 & 32 & A
\\
2018.3901 & 2062 & 13 & A
\\
2018.4256 & 721 & 25 & A
\\
2018.508 & $-$1102.3 & 53.5 & J
\\
2018.5540 & $-$1479 & 14 & A
\\
2018.5786 & $-$1626 & 14 & A
\\
2018.6087 & $-$1719 & 14 & A
\\
2018.6250 & $-$1764 & 18 & A
\\
2018.6277 & $-$1778 & 17 & A
\\
2018.628 & $-$1785.7 & 41.5 & J
\\
2018.6632 & $-$1809 & 28 & A
\\
2018.6633 & $-$1796 & 16 & A
\\
2018.6906 & $-$1818 & 17 & A
\\
2018.7070 & $-$1830 & 18 & A
\end{tabular}
\end{table}


\begin{table}[!h]
\caption{
Astrometric data from 1992.2 to 2003.6, the unit is mas.
}
\label{table:data-astrometry-1}
\centering
\begin{tabular}{lrlrlc}
date & $\ddec$ & $\uncertain{\dra}$ & $\dra$ & $\uncertain{\dra}$ & obs
\\
\hline
1992.224 & 170.2 & 3.8 & $-$9.9 & 3.7 & E
\\
1994.314 & 177.5 & 2.9 & $-$33.7 & 3.7 & E
\\
1995.4415 & 169.01 & 2.06 & $-$43.54 & 1.39 & A
\\
1995.534 & 170.3 & 3.5 & $-$41.8 & 3.0 & E
\\
1996.253 & 162.2 & 2.6 & $-$47.6 & 2.9 & E
\\
1996.427 & 160.2 & 4.4 & $-$50.3 & 1.7 & E
\\
1996.4853 & 155.18 & 3.53 & $-$53.31 & 3.22 & A
\\
1997.3676 & 140.64 & 1.41 & $-$58.81 & 1.11 & A
\\
1997.544 & 128.7 & 2.5 & $-$63.3 & 2.9 & E
\\
1998.373 & 120.4 & 2.5 & $-$69.2 & 3.5 & E
\\
1999.3347 & 96.92 & 0.55 & $-$68.26 & 0.77 & A
\\
1999.465 & 104.2 & 3.5 & $-$71.2 & 3.7 & E
\\
1999.5619 & 91.42 & 0.65 & $-$68.69 & 0.59 & A
\\
2000.3046 & 65.69 & 3.74 & $-$70.57 & 1.92 & A
\\
2000.3812 & 64.98 & 0.79 & $-$68.12 & 0.58 & A
\\
2000.472 & 62.0 & 2.4 & $-$58.7 & 4.1 & E
\\
2000.523 & 56.5 & 2.5 & $-$67.0 & 2.5 & E
\\
2000.5483 & 59.22 & 1.65 & $-$65.75 & 0.96 & A
\\
2000.7974 & 51.04 & 1.67 & $-$64.65 & 1.91 & A
\\
2001.3511 & 27.74 & 0.73 & $-$57.73 & 0.98 & A
\\
2001.502 & 22.0 & 1.6 & $-$52.4 & 3.2 & E
\\
2001.5729 & 17.48 & 0.78 & $-$52.75 & 1.01 & A
\\
2002.250 & $-$14.4 & 4.5 & $-$8.1 & 4.5 & E
\\
2002.335 & $-$9.7 & 3.0 & 4.0 & 3.0 & E
\\
2002.393 & $-$2.1 & 4.3 & 13.7 & 4.3 & E
\\
2002.409 & $-$0.1 & 3.7 & 15.5 & 3.7 & E
\\
2002.412 & 0.2 & 3.7 & 14.7 & 3.7 & E
\\
2002.414 & 1.0 & 3.7 & 14.7 & 3.7 & E
\\
2002.488 & 12.9 & 8.4 & 24.9 & 9.0 & E
\\
2002.578 & 18.5 & 3.7 & 28.1 & 3.7 & E
\\
2002.660 & 24.7 & 3.6 & 31.4 & 3.6 & E
\\
2003.1403 & 71.97 & 0.87 & 36.6 & 0.88 & A
\\
2003.214 & 64.5 & 0.4 & 38.6 & 0.4 & E
\\
2003.351 & 72.9 & 0.4 & 39.0 & 0.4 & E
\\
2003.356 & 72.7 & 0.4 & 38.3 & 0.4 & E
\\
2003.446 & 77.7 & 0.6 & 38.2 & 0.6 & E
\\
2003.451 & 78.3 & 0.5 & 38.9 & 0.5 & E
\\
2003.452 & 78.4 & 0.4 & 39.1 & 0.4 & E
\\
2003.454 & 79.7 & 0.4 & 38.9 & 0.4 & E
\\
2003.454 & 78.5 & 0.4 & 38.5 & 0.4 & E
\\
2003.550 & 83.2 & 0.4 & 38.6 & 0.4 & E
\\
2003.5565 & 84.14 & 0.54 & 36.52 & 0.70 & A
\end{tabular}
\end{table}

\begin{table}[!h]
\caption{
Astrometric data from 2003.6 to 2008.5, the unit is mas.
}
\label{table:data-astrometry-2}
\centering
\begin{tabular}{lrlrlc}
date & $\ddec$ & $\uncertain{\ddec}$ & $\dra$ & $\uncertain{\dra}$ & obs
\\
\hline
2003.676 & 89.7 & 0.4 & 38.3 & 0.4 & E
\\
2003.678 & 89.5 & 0.7 & 38.8 & 0.7 & E
\\
2003.6851 & 92.56 & 1.52 & 37.66 & 1.41 & A
\\
2003.761 & 94.7 & 0.5 & 37.7 & 0.5 & E
\\
2004.240 & 111.1 & 1.0 & 35.0 & 1.0 & E
\\
2004.325 & 114.1 & 0.3 & 34.7 & 0.3 & E
\\
2004.3265 & 115.23 & 0.50 & 32.99 & 0.43 & A
\\
2004.347 & 115.5 & 0.4 & 33.9 & 0.4 & E
\\
2004.443 & 118.3 & 0.4 & 33.9 & 0.4 & E
\\
2004.511 & 120.6 & 0.4 & 33.0 & 0.4 & E
\\
2004.513 & 121.0 & 0.4 & 33.2 & 0.4 & E
\\
2004.516 & 121.0 & 0.7 & 33.1 & 0.6 & E
\\
2004.516 & 121.0 & 0.5 & 33.4 & 0.5 & E
\\
2004.5647 & 123.0 & 0.62 & 31.12 & 0.6 & A
\\
2004.574 & 121.8 & 0.6 & 32.2 & 0.6 & E
\\
2004.574 & 122.7 & 0.4 & 32.2 & 0.4 & E
\\
2004.6605 & 125.65 & 0.74 & 30.37 & 0.85 & A
\\
2004.664 & 125.1 & 0.3 & 31.5 & 0.3 & E
\\
2004.670 & 125.3 & 0.5 & 31.9 & 0.5 & E
\\
2004.730 & 126.8 & 0.8 & 31.8 & 0.8 & E
\\
2004.730 & 127.2 & 0.3 & 31.1 & 0.3 & E
\\
2005.270 & 141.0 & 0.4 & 26.0 & 0.4 & E
\\
2005.3120 & 141.79 & 0.46 & 23.13 & 0.48 & A
\\
2005.366 & 143.1 & 0.4 & 25.1 & 0.4 & E
\\
2005.371 & 143.3 & 0.4 & 24.7 & 0.4 & E
\\
2005.374 & 143.3 & 0.5 & 24.7 & 0.5 & E
\\
2005.467 & 144.9 & 0.4 & 24.3 & 0.4 & E
\\
2005.4962 & 146.84 & 0.47 & 22.09 & 0.26 & A
\\
2005.5674 & 147.93 & 1.30 & 21.23 & 1.16 & A
\\
2005.570 & 146.9 & 0.4 & 23.5 & 0.4 & E
\\
2005.576 & 147.3 & 0.4 & 23.0 & 0.4 & E
\\
2005.5811 & 148.36 & 0.24 & 21.31 & 0.2 & A
\\
2006.324 & 159.7 & 0.7 & 15.7 & 0.9 & E
\\
2007.545 & 173.6 & 0.8 & 1.3 & 1.0 & E
\\
2007.550 & 173.1 & 0.5 & 2.6 & 0.5 & E
\\
2007.686 & 173.9 & 0.6 & 1.0 & 0.6 & E
\\
2007.687 & 173.9 & 0.6 & 0.4 & 0.6 & E
\\
2008.148 & 177.0 & 0.4 & $-$6.0 & 0.4 & E
\\
2008.197 & 177.0 & 0.4 & $-$6.5 & 0.4 & E
\\
2008.268 & 178.0 & 0.4 & $-$7.4 & 0.4 & E
\\
2008.3703 & 179.68 & 0.16 & $-$10.27 & 0.16 & A
\\
2008.456 & 178.2 & 0.3 & $-$9.7 & 0.4 & E
\end{tabular}
\end{table}

\begin{table}[!h]
\caption{
Astrometric data from 2008.5 to 2010,7, the unit is mas.
}
\label{table:data-astrometry-3}
\centering
\begin{tabular}{lrlrlc}
date & $\ddec$ & $\uncertain{\ddec}$ & $\dra$ & $\uncertain{\dra}$ & obs
\\
\hline
2008.472 & 178.6 & 0.4 & $-$9.4 & 0.4 & E
\\
2008.473 & 178.2 & 0.5 & $-$9.0 & 0.5 & E
\\
2008.5619 & 180.25 & 0.16 & $-$12.38 & 0.16 & A
\\
2008.593 & 177.6 & 1.4 & $-$10.0 & 1.6 & E
\\
2008.601 & 178.2 & 0.4 & $-$11.8 & 0.4 & E
\\
2008.708 & 179.2 & 0.4 & $-$12.6 & 0.4 & E
\\
2009.185 & 179.1 & 0.8 & $-$18.4 & 0.8 & E
\\
2009.273 & 179.2 & 0.4 & $-$19.1 & 0.4 & E
\\
2009.300 & 179.3 & 0.4 & $-$19.6 & 0.4 & E
\\
2009.303 & 179.6 & 0.4 & $-$19.3 & 0.4 & E
\\
2009.336 & 179.2 & 0.4 & $-$19.5 & 0.4 & E
\\
2009.336 & 179.2 & 0.4 & $-$19.4 & 0.4 & E
\\
2009.3402 & 180.75 & 0.16 & $-$21.02 & 0.15 & A
\\
2009.371 & 179.0 & 0.4 & $-$19.7 & 0.4 & E
\\
2009.502 & 179.0 & 0.5 & $-$20.9 & 0.5 & E
\\
2009.505 & 179.2 & 0.4 & $-$21.2 & 0.4 & E
\\
2009.557 & 179.4 & 0.4 & $-$21.3 & 0.4 & E
\\
2009.557 & 179.5 & 0.4 & $-$21.9 & 0.4 & E
\\
2009.5619 & 180.53 & 0.20 & $-$23.34 & 0.16 & A
\\
2009.606 & 179.5 & 0.4 & $-$22.3 & 0.4 & E
\\
2009.6906 & 180.33 & 0.18 & $-$24.78 & 0.18 & A
\\
2009.718 & 179.1 & 0.4 & $-$23.9 & 0.4 & E
\\
2009.776 & 179.0 & 0.4 & $-$24.1 & 0.4 & E
\\
2010.234 & 177.5 & 0.4 & $-$29.5 & 0.4 & E
\\
2010.239 & 177.1 & 0.4 & $-$29.0 & 0.4 & E
\\
2010.239 & 176.6 & 0.4 & $-$29.6 & 0.4 & E
\\
2010.245 & 176.8 & 0.4 & $-$29.3 & 0.4 & E
\\
2010.3429 & 177.88 & 0.17 & $-$31.87 & 0.14 & A
\\
2010.351 & 176.6 & 0.4 & $-$30.6 & 0.4 & E
\\
2010.444 & 176.1 & 0.4 & $-$31.8 & 0.4 & E
\\
2010.455 & 175.4 & 0.5 & $-$31.2 & 0.5 & E
\\
2010.455 & 176.1 & 0.4 & $-$31.4 & 0.4 & E
\\
2010.455 & 175.9 & 0.4 & $-$31.8 & 0.4 & E
\\
2010.46 & 176.1 & 0.6 & $-$31.3 & 0.6 & E
\\
2010.5127 & 177.15 & 0.15 & $-$33.67 & 0.15 & A
\\
2010.616 & 174.4 & 0.4 & $-$31.8 & 0.4 & E
\\
2010.619 & 175.1 & 0.4 & $-$33.9 & 0.4 & E
\\
2010.622 & 175.1 & 0.4 & $-$33.3 & 0.4 & E
\\
2010.6222 & 176.61 & 0.16 & $-$34.82 & 0.17 & A
\\
2010.624 & 175.3 & 0.4 & $-$33.2 & 0.4 & E
\\
2010.627 & 174.9 & 0.4 & $-$33.5 & 0.4 & E
\\
2010.676 & 174.1 & 0.6 & $-$33.6 & 0.6 & E
\\
2010.679 & 174.4 & 0.4 & $-$34.2 & 0.4 & E
\end{tabular}
\end{table}

\begin{table}[!h]
\caption{
Astrometric data from 2011.2 to 2013.7, the unit is mas.
}
\label{table:data-astrometry-4}
\centering
\begin{tabular}{lrlrlc}
date & $\ddec$ & $\uncertain{\ddec}$ & $\dra$ & $\uncertain{\dra}$ & obs
\\
\hline
2011.238 & 170.7 & 0.4 & $-$39.7 & 0.4 & E
\\
2011.241 & 170.6 & 1.9 & $-$39.7 & 1.9 & E
\\
2011.244 & 170.4 & 0.4 & $-$39.1 & 0.4 & E
\\
2011.249 & 170.3 & 0.5 & $-$39.5 & 0.5 & E
\\
2011.312 & 169.8 & 0.4 & $-$40.3 & 0.4 & E
\\
2011.313 & 169.9 & 0.5 & $-$38.9 & 0.5 & E
\\
2011.315 & 169.7 & 0.4 & $-$40.5 & 0.4 & E
\\
2011.337 & 169.3 & 0.3 & $-$40.4 & 0.3 & E
\\
2011.4031 & 170.75 & 0.21 & $-$42.62 & 0.21 & A
\\
2011.443 & 168.4 & 0.4 & $-$41.4 & 0.4 & E
\\
2011.5455 & 169.01 & 0.15 & $-$44.08 & 0.15 & A
\\
2011.553 & 167.6 & 0.6 & $-$42.5 & 0.6 & E
\\
2011.613 & 166.9 & 0.7 & $-$42.7 & 0.7 & E
\\
2011.6441 & 167.8 & 0.28 & $-$44.93 & 0.18 & A
\\
2011.689 & 166.3 & 0.7 & $-$44.0 & 0.7 & E
\\
2011.695 & 166.2 & 1.0 & $-$42.8 & 1.0 & E
\\
2011.695 & 166.1 & 0.4 & $-$43.9 & 0.4 & E
\\
2011.698 & 166.0 & 0.4 & $-$43.9 & 0.4 & E
\\
2011.722 & 165.6 & 0.4 & $-$44.1 & 0.4 & E
\\
2012.202 & 159.8 & 0.7 & $-$48.0 & 0.7 & E
\\
2012.339 & 158.4 & 0.4 & $-$49.6 & 0.4 & E
\\
2012.3703 & 159.94 & 0.15 & $-$51.21 & 0.14 & A
\\
2012.497 & 156.2 & 0.4 & $-$50.1 & 0.4 & E
\\
2012.533 & 156.2 & 0.4 & $-$51.4 & 0.4 & E
\\
2012.544 & 156.1 & 0.4 & $-$51.1 & 0.4 & E
\\
2012.552 & 155.8 & 0.4 & $-$51.7 & 0.4 & E
\\
2012.552 & 155.8 & 0.4 & $-$51.7 & 0.4 & E
\\
2012.552 & 155.8 & 0.4 & $-$51.7 & 0.4 & E
\\
2012.5619 & 157.13 & 0.26 & $-$53.02 & 0.18 & A
\\
2012.700 & 153.4 & 0.4 & $-$52.0 & 0.4 & E
\\
2013.161 & 147.1 & 0.4 & $-$55.3 & 0.4 & E
\\
2013.240 & 145.8 & 0.4 & $-$55.5 & 0.4 & E
\\
2013.317 & 144.6 & 0.4 & $-$56.5 & 0.4 & E
\\
2013.366 & 143.4 & 0.4 & $-$56.7 & 0.4 & E
\\
2013.420 & 142.7 & 0.5 & $-$56.7 & 0.5 & E
\\
2013.437 & 142.7 & 0.7 & $-$56.0 & 0.7 & E
\\
2013.494 & 140.9 & 0.5 & $-$57.3 & 0.5 & E
\\
2013.502 & 141.1 & 0.4 & $-$57.3 & 0.4 & E
\\
2013.587 & 139.5 & 0.4 & $-$57.9 & 0.4 & E
\\
2013.590 & 139.3 & 1.1 & $-$57.3 & 1.1 & E
\\
2013.617 & 139.0 & 0.6 & $-$58.3 & 0.6 & E
\end{tabular}
\end{table}

\begin{table}[!h]
\caption{
Astrometric data from 2015.4 to 2018.7, the unit is mas.
}
\label{table:data-astrometry-5}
\centering
\begin{tabular}{lrlrlc}
date & $\ddec$ & $\uncertain{\ddec}$ & $\dra$ & $\uncertain{\dra}$ & obs
\\
\hline
2015.432 & 97.7 & 0.4 & $-$66.2 & 0.4 & E
\\
2015.517 & 95.3 & 0.4 & $-$66.7 & 0.4 & E
\\
2015.706 & 90.1 & 0.5 & $-$67.0 & 0.5 & E
\\
2015.747 & 88.5 & 0.4 & $-$66.8 & 0.4 & E
\\
2016.221 & 73.3 & 0.4 & $-$66.1 & 0.4 & E
\\
2016.287 & 70.8 & 0.5 & $-$66.0 & 0.5 & E
\\
2016.325 & 69.7 & 0.8 & $-$66.0 & 0.8 & E
\\
2016.3374 & 71.08 & 0.25 & $-$67.14 & 0.17 & A
\\
2016.369 & 68.7 & 0.8 & $-$66.2 & 0.8 & E
\\
2016.525 & 62.7 & 0.5 & $-$64.4 & 0.5 & E
\\
2016.530 & 62.7 & 0.5 & $-$65.1 & 0.5 & E
\\
2016.5318 & 64.49 & 0.20 & $-$66.21 & 0.21 & A
\\
2017.3429 & 32.89 & 0.23 & $-$57.42 & 0.17 & A
\\
2017.3484 & 33.1 & 0.33 & $-$57.17 & 0.25 & A
\\
2017.6167 & 20.61 & 0.27 & $-$51.33 & 0.27 & A
\\
2017.6496 & 19.22 & 0.32 & $-$50.22 & 0.25 & A
\\
2018.2088 & $-$8.20 & 0.31 & $-$22.14 & 0.30 & A
\\
2018.2225 & $-$8.27 & 0.40 & $-$20.64 & 0.49 & A
\\
2018.2444 & $-$9.18 & 0.31 & $-$18.20 & 0.33 & A
\\
2018.3812 & $-$6.73 & 0.98 & 1.07 & 1.12 & A
\\
2018.3949 & $-$6.41 & 0.47 & 4.38 & 0.84 & A
\\
2018.6742 & 24.86 & 0.47 & 27.66 & 0.39 & A
\end{tabular}
\end{table}

\end{document}